\RequirePackage{amsmath}
\documentclass[orivec]{llncs}

\usepackage{chngcntr}
\usepackage{wasysym}
\usepackage{wrapfig}
\usepackage{sectsty}
\usepackage{graphicx,amssymb,calc}
\usepackage{pifont}
\usepackage{etoolbox}
\usepackage{thmtools}
\usepackage{accents}
\usepackage[utf8]{inputenc}
\usepackage{multirow}
\usepackage{hhline}
\usepackage{minibox}
\usepackage{xparse}
\usepackage[f]{esvect}  
\usepackage{bm}
\usepackage{mathtools}
\usepackage{changepage}
\usepackage{stmaryrd}
\usepackage{titlesec}

\usepackage{algorithm}
\usepackage[noend]{algpseudocode}
\usepackage{bussproofs,setspace,array}
\usepackage{lscape}
\usepackage{caption}
\usepackage{wasysym}
\usepackage{color}   
\usepackage[usenames,dvipsnames]{xcolor}
\usepackage{xkvltxp}
\usepackage{parskip} 
\usepackage{longtable}
\usepackage{url}
\usepackage{xcolor}
\usepackage{hyperref}
\hypersetup{ 
  colorlinks,                    
  linkcolor={red!50!black},      
  citecolor={blue},              
  urlcolor={blue!80!black},      
  backref=page                   
}
\usepackage{cleveref}
\usepackage{nameref}
\usepackage{authblk}
\usepackage{tikz}
\usetikzlibrary{calc,shapes.multipart,chains,arrows,positioning}
\usetikzlibrary{decorations.pathreplacing}

\definecolor{timberwolf}{rgb}{0.86, 0.84, 0.82}
\definecolor{shadow}{rgb}{0.54, 0.47, 0.36}
\definecolor{sepia}{rgb}{0.44, 0.26, 0.08}
\definecolor{sanddune}{rgb}{0.59, 0.44, 0.09}
\definecolor{pastelbrown}{rgb}{0.51, 0.41, 0.33}
\definecolor{goldenbrown}{rgb}{0.6, 0.4, 0.08}
\definecolor{palebrown}{rgb}{0.6, 0.46, 0.33}

\newcommand{\code}[1]{\ensuremath{\mathtt{#1}}}

\newcommand{\hformn}[2]{\ensuremath{{#1}{\mapsto}{#2}}}

\newcommand{\hformnt}[3]{\ensuremath{{#2}{\xmapsto{#1}}{#3}}}
\newcommand{\hformntShort}[3]{\ensuremath{{#2}{\overset{\mkern-2mu\hspace{-1pt}#1}{\scalebox{1.3}[1]{$\mapsto$}}}{#3}}}
\newcommand{\hformp}[2]{\ensuremath{{#1}{(#2)}}}        
\newcommand{\pred}[1]{\ensuremath{\mathsf{#1}}}
\newcommand{\proc}[1]{\ensuremath{\mathsf{#1}}}

\newcommand{\evalVar}[2]{\ensuremath{{#2}({#1})}}
\newcommand{\evalForm}[2]{\ensuremath{\llbracket {#1} \rrbracket_{#2}}}
\newcommand{\modelExtOne}[3]{\ensuremath{[{#1}|#2{:}#3]}}

\newcommand{\setenum}[1]{\{\, #1 \,\}}

\def\entails{\mathrel{
  \raisebox{0.1em}{\scalebox{1}[0.85]{\ensuremath{|}}}
  \mkern-3mu\mkern-1mu
  \raisebox{0.08em}{\scalebox{0.95}[1]{\ensuremath{-}}}}}

\def\satisfies{\mathrel{
  \raisebox{0.1em}{\scalebox{1}[0.85]{\ensuremath{|}}}
  \mkern-2.7mu\mkern-1mu
  \raisebox{0.08em}{\scalebox{0.95}[1]{\ensuremath{=}}}}}
\def\nsatisfies{\mathrel{
  \raisebox{0.085em}{\scalebox{0.92}[0.92]{\ensuremath{\not}}}
  \mkern-4mu\mkern-1mu
  \satisfies}}

\newcommand{\satentail}[2]{#1 \satisfies #2}

\def\synequiv{\mathbin{\cong}}

\def\ltmodel{\mathbin{\prec}}

\newcommand{\val}[1]{\ensuremath{\mathsf{#1}}}                

\newcommand{\cpointer}[1]{   
  {\raisebox{2.2pt}{\footnotesize $\ast$}}\,#1}

\def\shorteq{\mathbin{\ensuremath{\scalebox{0.75}[1]{$=$}}}}  
\def\shortneq{\mathbin{\ensuremath{{!}\shorteq}}}  
\def\shorteqeq{\mathbin{{\shorteq}{\shorteq}}}  

\def\hunions{\circ}  
\def\disjoins{\mathrel{\#}}   

\def\mult/{\ensuremath{\hspace{0.1em}{\cdot}\hspace{0.1em}}}

\def\defBNF/{\Coloneqq}

\makeatletter
\newcommand{\algComment}[1]{{\color{Brown}{//\,\textit{#1}}}}
\newcommand{\algCommentColor}[2]{{\color{#1}{//\,\textit{#2}}}}
\algnewcommand{\LineComment}[1]{\Statex \hskip\ALG@thistlm \algComment{#1}}
\algnewcommand{\LineCommentIndent}[1]{\Statex \hskip\ALG@thistlm \hskip\algorithmicindent \algComment{#1}}
\algnewcommand{\CommentColor}[2]{\hfill \algCommentColor{#1}{#2}}
\algnewcommand{\CommentLn}[1]{\Statex \hskip\ALG@thistlm \algComment{#1}}
\algnewcommand{\CommentLnColor}[2]{\Statex \hskip\ALG@thistlm \algCommentColor{#1}{#2}}
\algnewcommand{\CommentLnIndent}[1]{\Statex \hskip\ALG@thistlm \hskip\algorithmicindent \algComment{#1}}
\algnewcommand{\CommentLnIndentColor}[2]{\Statex \hskip\ALG@thistlm \hskip\algorithmicindent \algCommentColor{#1}{#2}}

\algrenewcommand{\algorithmiccomment}[1]{\hfill \algComment{#1}}
\algrenewcommand\textproc{}

\algrenewcommand\alglinenumber[1]{\hspace{-2em}\scriptsize #1:}
\makeatother

\algrenewcommand\algorithmicindent{2em}  

\newcommand{\psCall}[2]{{{\rm {#1}}}\ensuremath{(#2)}}             
\newcommand{\psKey}[1]{\textbf{#1}}                             
\newcommand{\psAssign}[2]{\ensuremath{#1} \ensuremath{\leftarrow} \ensuremath{#2}}    
\newcommand{\psReturn}[1]{\psKey{return} \ensuremath{#1}}

\newcommand{\cmark}{\text{\ding{51}}}%
\newcommand{\xmark}{\text{\ding{55}}}%

\newcommand{\rulename}[1]{$(#1)$}

\newcommand{\rulesidecondright}[1]{    
  {\fontsize{8pt}{8pt}\selectfont{{$#1$}}}}

\def\mtor{\mathrel{\pmb{\vee}}}
\def\mtand{\mathrel{\pmb{\wedge}}}
\def\mtimply{\mathrel{\rightarrow}}

\def\inserttrace{\mathbin{::}}
\def\concattrace{\mathbin{\textrm{@}}}

\def\membertrace{\mathbin{\in}}
\def\nmembertrace{\mathbin{\not\in}}

\def\qedhere{$\hfill\boxempty$}

\newcounter{myequationNo}

\sectionfont{\fontsize{12}{15}\selectfont}  
\subsectionfont{\fontsize{11}{12}\selectfont}  

\newcommand{\sburl}{\textsf{http://loris{-}5.d2.comp.nus.edu.sg/songbird/}}

\newcommand{\xmarksb}{\mathord{\xmark_{\text{sb}}}}
\newcommand{\xmarkother}{\mathord{\xmark_{\text{o}}}}
\newcommand{\cmarksb}{\mathord{\cmark_{\joinrel\joinrel\/ \text{sb}}}}
\newcommand{\cmarkother}{\mathord{\cmark_{\joinrel\joinrel\/ \text{o}}}}

\def\valNil{\mathord{\val{nil}}}
\def\valValid{\mathrel{\mathsf{True}}}
\def\valUnknown{\mathrel{\mathsf{False}}}
\def\statusValid{\mathord{\boldsymbol{\checkmark}}}
\def\statusUnknown{\mathrel{\boldsymbol{?}}}

\def\valTrue{\mathord{\mathit{true}}}

\def\predEmp{\mathord{\pred{emp}}}

\def\predP{\mathord{\pred{P}}}
\def\formF{\mathord{F}}

\def\predPi{\mathord{\predP_i}}

\def\predLs{\mathord{\pred{ls}}}

\def\predLsEven{\mathord{\pred{lsEven}}}

\def\predTmp{\mathord{\mathsf{tmp}}}

\def\sepAnte{\mathord{,\,\,}}

\def\ruleR{\mathord{R}}
\def\ruleRs{\mathord{\mathcal{S}}}

\def\ruleHypo{\mathord{\mathsf{AH}}}

\def\ruleInduction{\mathord{\mathsf{Ind}}}
\def\rulePureEntail{\mathord{\entails_{\text{pure}}}}
\def\ruleFalseLeftOne{\mathord{\bot\,\mathsf{L_1}}}
\def\ruleFalseLeftTwo{\mathord{\bot\,\mathsf{L_2}}}

\def\ruleStarData{\mathord{*\,{\mapsto}}}
\def\ruleStarPred{\mathord{*\,{\predP}}}
\def\ruleEmpLeft{\mathord{\predEmp\mathsf{L}}}
\def\ruleEmpRight{\mathord{\predEmp\mathsf{R}}}

\def\ruleExistsLeft{\mathord{\exists\,\mathsf{L}}}
\def\ruleExistsRight{\mathord{\exists\,\mathsf{R}}}

\def\rulePredIntroRight{\mathord{\predP\mathsf{R}}}
\def\ruleEqualLeft{\mathord{=\joinrel\mathsf{L}}}
\def\ruleEqualRight{\mathord{=\joinrel\mathsf{R}}}

\def\setInt{\mathord{\code{Int}}}
\def\setNats{\mathord{\mathbb{N}}}

\def\setSort{\mathord{\code{Sort}}}
\def\setVar{\mathord{\code{Var}}}
\def\setVal{\mathord{\code{Val}}}
\def\setLoc{\mathord{\code{Loc}}}

\def\setempty{\mathord{\varnothing}}

\def\theoryLA{\mathord{\mathrm{LA}}}

\def\theorySLSH{\mathord{\mathrm{SL}_{\mathsf{ID}}}}

\newcommand{\freevars}[1]{\ensuremath{\mathtt{FV}(#1)}}

\newcommand{\funcDom}[1]{{\ensuremath{\mathrm{dom}(#1)}}}
\newcommand{\cardFunc}[1]{{\ensuremath{|#1|}}}

\def\varRule{\mathord{R}}
\def\varRuleSet{\mathord{\mathcal{R}}}
\def\varRuleSelected{\mathord{\mathcal{S}}}
\def\varTree{\mathcal{T}}

\def\varSort{\iota}

\def\varTree{\mathord{\mathcal{T}}}

\def\varHypo{\mathord{\mathcal{H}}}

\def\varTrace{\mathord{\rho}}
\def\varTracePrim{\mathord{\mathcal{\rho}'}}

\def\procProve{\proc{Prove}}

\def\songbird{\mathord{\textsf{Songbird}}}
\def\songbirdSI{\mathord{\mathsf{Songbird_{\text{SI}}}}}

\def\slide{\textsf{Slide}}
\def\spen{\textsf{Spen}}
\def\sleek{\textsf{Sleek}}
\def\cyclist{\textsf{Cyclist}}

\newcommand\minipagePureEntail[1]{
\begin{tabular}{c}
\begin{minipage}{#1}
\raggedleft
\begin{prooftree}
  \def\ScoreOverhang{0em}
  \AxiomC{}
  \def\extraVskip{3pt}
  \LeftLabel{\rulename{\rulePureEntail}}
  \RightLabel{\rulesidecondright{\Pi_1 \,{\Rightarrow}\, \Pi_2}}
  \UnaryInfC{$
    \varHypo \sepAnte \varTrace \sepAnte
    \Pi_1
    \,{\entails}\,
    \Pi_2
  $}
\end{prooftree}
\end{minipage}
\end{tabular}}

\newcommand\minipageFalseLeftOne[1]{
\begin{tabular}{c}
\begin{minipage}{#1}
\begin{prooftree}
  \def\ScoreOverhang{-0.1em}
  \AxiomC{$~$}
  \LeftLabel{\rulename{\ruleFalseLeftOne}\hspace{-0.1em}}
  \def\extraVskip{3pt}
  \UnaryInfC{$
    \varHypo \sepAnte \varTrace \sepAnte
    F_1 \,{\wedge}\, u{\neq}u
    \,{\entails}\,
    F_2
  $}
\end{prooftree}
\end{minipage}
\end{tabular}}

\newcommand\minipageFalseLeftTwo[1]{
\begin{tabular}{c}
\begin{minipage}{#1}
\begin{prooftree}
  \def\ScoreOverhang{0em}
  \AxiomC{$~$}
  \LeftLabel{\rulename{\ruleFalseLeftTwo}}
  \def\extraVskip{3pt}
  \UnaryInfC{$
    \varHypo \sepAnte \varTrace \sepAnte
    F_1 \,{*}\, \hformntShort{\varSort_1}{u}{\vec{v}} \,{*}\,
    \hformntShort{\varSort_2}{u}{\vec{w}}
    \,{\entails}\,
    F_2
  $}
\end{prooftree}
\end{minipage}
\end{tabular}}

\newcommand\minipageEqualLeft[1]{
\begin{tabular}{c}
\begin{minipage}{#1}
\raggedleft
\begin{prooftree}
  \def\ScoreOverhang{0em}
  \AxiomC{$
    \varHypo \sepAnte\, \varTracePrim \sepAnte\,
    F_1[u/v]
    \entails
    F_2 [u/v]
  $}
  \def\extraVskip{3pt}
  \LeftLabel{\rulename{\ruleEqualLeft}}
  \UnaryInfC{$
    \varHypo \sepAnte\, \varTrace \sepAnte\,
    F_1 \wedge u{=}v
    \entails
    F_2
  $}
\end{prooftree}
\end{minipage}
\end{tabular}}

\newcommand\minipageEqualRight[1]{
\begin{tabular}{c}
\begin{minipage}{#1}
\raggedleft
\begin{prooftree}
  \def\ScoreOverhang{0em}
  \AxiomC{$
    \varHypo \sepAnte\, \varTracePrim \sepAnte\,
    F_1
    \entails
    \exists \vec{x}.F_2
  $}
  \def\extraVskip{3pt}
  \LeftLabel{\rulename{\ruleEqualRight}\hspace{-0.1em}}
  \UnaryInfC{$
    \varHypo \sepAnte\, \varTrace \sepAnte\,
    F_1
    \entails
    \exists \vec{x}.(F_2 \wedge u{=}u)
  $}
\end{prooftree}
\end{minipage}
\end{tabular}}

\newcommand\minipageExistsLeft[1]{
\begin{tabular}{c}
\begin{minipage}{#1}
\raggedleft
\begin{prooftree}
  \def\ScoreOverhang{-0.1em}
  \AxiomC{$
    \varHypo \sepAnte \varTracePrim \sepAnte
    F_1 [u/x]
    \,{\entails}\,
    F_2
  $}
  \def\extraVskip{3pt}
  \LeftLabel{\rulename{\ruleExistsLeft}\hspace{-0.1em}}
  \RightLabel{\hspace{-0.3em}\rulesidecondright{u \,{\not\in}\, \freevars{F_2}}}
  \UnaryInfC{$
    \varHypo \sepAnte \varTrace \sepAnte
    \exists x. F_1
    \,{\entails}\,
    F_2
  $}
\end{prooftree}
\end{minipage}
\end{tabular}}

\newcommand\minipageExistsRight[1]{
\begin{tabular}{c}
\begin{minipage}{#1}
\raggedleft
\begin{prooftree}
  \def\ScoreOverhang{-0.1em}
  \AxiomC{$
    \varHypo \sepAnte \varTracePrim \sepAnte
    F_1
    \,{\entails}\,
    F_2 [e/x]
  $}
  \def\extraVskip{3pt}
  \LeftLabel{\rulename{\ruleExistsRight}}
  \UnaryInfC{$
    \varHypo \sepAnte \varTrace \sepAnte
    F_1
    \,{\entails}\,
    \exists x.F_2
  $}
\end{prooftree}
\end{minipage}
\end{tabular}}

\newcommand\minipageEmpLeftOne[1]{
\begin{tabular}{c}
\begin{minipage}{#1}
\begin{prooftree}
  \def\ScoreOverhang{0em}
  \AxiomC{$
    \varHypo \sepAnte\, \varTracePrim \sepAnte\,
    F_1
    \entails
    F_2
  $}
  \def\extraVskip{2pt}
  \LeftLabel{\rulename{\ruleEmpLeft}}
  \UnaryInfC{$
    \varHypo \sepAnte\, \varTrace \sepAnte\,
    F_1 * \predEmp
    \entails
    F_2
  $}
\end{prooftree}
\end{minipage}
\end{tabular}}

\newcommand\minipageEmpRightOne[1]{
\begin{tabular}{c}
\begin{minipage}{#1}
\begin{prooftree}
  \def\ScoreOverhang{0em}
  \AxiomC{$
    \varHypo \sepAnte\, \varTracePrim \sepAnte\,
    F_1
    \entails
    \exists \vec{x}. F_2
  $}
  \def\extraVskip{2pt}
  \LeftLabel{\rulename{\ruleEmpRight}}
  \UnaryInfC{$
    \varHypo \sepAnte\, \varTrace \sepAnte\,
    F_1
    \entails
    \exists \vec{x}. (F_2 * \predEmp)
  $}
\end{prooftree}
\end{minipage}
\end{tabular}}

\newcommand\minipageStarData[1]{
\begin{tabular}{c}
\begin{minipage}{#1}
\begin{prooftree}
  \def\ScoreOverhang{0em}
  \AxiomC{$
    \varHypo \sepAnte\, \varTracePrim \sepAnte\,
    F_1
    \entails
    \exists \vec{x}.
    (F_2 \wedge u{=}t \wedge \vec{v}{=}\vec{w})
  $}
  \def\extraVskip{2pt}
  \LeftLabel{\rulename{\ruleStarData}}
  \RightLabel{\rulesidecondright{
    (u, \vec{v}) \,{\disjoins}\, \vec{x}
  }}
  \UnaryInfC{$
    \varHypo \sepAnte\, \varTrace \sepAnte\,
    F_1 * \hformntShort{\varSort}{u}{\vec{v}}
    \entails
    \exists \vec{x}.(F_2 * \hformntShort{\varSort}{t}{\vec{w}})
  $}
\end{prooftree}
\end{minipage}
\end{tabular}}

\newcommand\minipageStarPred[1]{
\begin{tabular}{c}
\begin{minipage}{#1}
\begin{prooftree}
  \def\ScoreOverhang{0em}
  \AxiomC{$
    \varHypo \sepAnte\, \varTracePrim \sepAnte\,
    F_1
    \entails
    \exists \vec{x}. (F_2 \wedge \vec{u}{=}\vec{v})
  $}
  \def\extraVskip{2pt}
  \LeftLabel{\rulename{\ruleStarPred}}
  \RightLabel{\rulesidecondright{
      \vec{u} \,{\disjoins}\, \vec{x}
  }}
  \UnaryInfC{$
    \varHypo \sepAnte\, \varTrace \sepAnte\,
    F_1 * \hformp{\predP}{\vec{u}}
    \entails
    \exists \vec{x}. (F_2 * \hformp{\predP}{\vec{v}})
  $}
\end{prooftree}
\end{minipage}
\end{tabular}}

\newcommand\minipagePredIntroRight[1]{
\begin{tabular}{c}
\begin{minipage}{#1}
\begin{prooftree}
  \def\ScoreOverhang{0em}
  \AxiomC{$
    \varHypo \sepAnte\, \varTracePrim \sepAnte\,
    F_1 \entails
    \exists \vec{x}.
    (F_2 * \hformp{\formF^{\predP}_i}{\vec{u}})
  $}
  \def\extraVskip{3pt}
  \LeftLabel{\rulename{\rulePredIntroRight}}
  \RightLabel{\parbox{8em}{
    \rulesidecondright{
    \hformp{\formF^{\predP}_i}{\vec{u}} \text{~is~one~of~the}}\\
    \rulesidecondright{
    \text{definition~cases~of~} \hformp{\predP}{\vec{u}}}
  }}
  \UnaryInfC{$
    \varHypo \sepAnte\, \varTrace \sepAnte\,
    F_1 \entails
    \exists \vec{x}.(F_2 * \hformp{\predP}{\vec{u}})
  $}
\end{prooftree}
\end{minipage}
\end{tabular}}

\titlespacing*{\section}{0pt}{1ex plus .1ex}{0ex plus .1ex}
\titlespacing*{\subsection}{0pt}{1ex plus .1ex}{0.5ex plus .1ex}
\titlespacing{\paragraph}{0pt}{1em}{1em}

\begin{document}

\pagestyle{plain}  

\title{Automated Mutual Explicit Induction Proof in Separation Logic}

\author{
  Quang-Trung Ta  \quad
  Ton Chanh Le  \quad
  Siau-Cheng Khoo  \quad
  Wei-Ngan Chin
  \\ \vspace{-0.8em}
  \textup{\footnotesize
    \{taqt, chanhle, khoosc, chinwn\}@comp.nus.edu.sg
  }}

\institute{{\normalsize School of Computing, National University of Singapore}}

\maketitle


\begin{abstract}

  We present a sequent-based deductive system for automatically proving
  entailments in separation logic by using mathematical induction. Our
  technique, called {\em mutual explicit induction proof}, is an instance of
  Noetherian induction. Specifically, we propose a novel induction principle
  on a well-founded relation of separation logic model, and follow the {\em
    explicit} induction methods to implement this principle as inference rules,
  so that it can be easily integrated into a deductive system. We also
  support {\em mutual} induction, a natural feature of implicit induction,
  where the goal entailment and other entailments derived during the proof search
  can be used as hypotheses to prove each other. We have implemented a
  prototype prover and evaluated it on a benchmark of handcrafted entailments
  as well as benchmarks from a separation logic competition.

\end{abstract}


\section{Introduction}

Separation logic (SL)~\cite{OHearnRY01,Reynolds02} has been actively used
recently to reason about imperative programs that alter data structures.
For example, the static analysis tool Infer~\cite{FacebookInfer} of
Facebook has been using SL to discover critical memory safety bugs in
Android and iOS applications. One of the pivotal features making the
success of SL is the {\em separating conjunction} operator $(*)$, which is
used to describe the separation of computer memory. In particular, the
assertion $p * q$ denotes a memory portion which can be decomposed into two
{\em disjoint\/} sub-portions held by $p$ and $q$, respectively. In
addition, SL is also equipped with the ability for users to define
inductive heap predicates \cite{Reynolds00,BrotherstonDP11,IosifRS13}. The
combination of the separating conjunction and inductive heap predicates
makes SL expressive enough to model various types of recursive data
structures, such as linked lists and trees.

However, this powerful expressiveness also poses challenges in reasoning
about SL entailments. Considerable
researches have been conducted on the SL entailment proving problem,
including the works \cite{BrotherstonDP11,BrotherstonGP12,ChuJT15} related
to mathematical induction. In particular, Brotherston et al.
~\cite{BrotherstonDP11,BrotherstonGP12} propose the {\em cyclic proof},
which allows proof trees to contain cycles, and can be perceived as
infinite derivation trees. Furthermore, during the proof derivation, induction
hypotheses are not explicitly identified via applications of induction
rules; instead, they are implicitly obtained via the discovery of valid
cycle proofs. Consequently, a soundness condition needs to be checked
globally on proof trees. On the other hand, Chu et al.~\cite{ChuJT15} apply
{\em structural induction\/} on inductive heap predicates for proving SL
entailments. During proof search, this technique dynamically uses derived
entailments as induction hypotheses. When applying induction hypotheses, it
performs a local check to ensure that predicates in the target entailments
are substructures of predicates in the entailments captured as hypotheses.
This dynamicity in hypothesis generation enables multiple induction
hypotheses within a single proof path to be exploited; however, it does not
admit hypotheses obtained from different proof paths.

In this work, we develop a sequent-based deductive system for proving SL
entailments by using mathematical induction. Our technique is an instance
of Noetherian induction ~\cite{Bundy01}, where we propose a novel induction
principle based on a well-founded relation of SL models. Generally, proof
techniques based on Noetherian induction are often classified into two
categories, i.e., {\em explicit} and {\em implicit}
induction ~\cite{Bundy01}, and each of them presents advantages over the
other. We follow the explicit induction methods to implement the induction
principle as inference rules, so that it can be easily integrated into a
deductive system, and the soundness condition can be checked locally in
each application of inference rules. In addition, since the well-founded relation defined
in our induction principle does not depend directly on the substructure
relationship, induction hypotheses gathered in one proof path can be used
for hypothesis applications at other proof paths of the entire proof
tree. Thus, our induction principle also favors {\em mutual
induction}, a natural feature of {\em implicit induction}, in which the
goal entailment and other entailments derived during the proof search can be
used as hypotheses to prove each other. Our proof technique, therefore,
does not restrict induction hypotheses to be collected from only one proof
path, but rather from all derived paths of the proof tree.

{\bf Related work.} The entailment proving problem in SL has been actively
studied recently. Various sound and complete techniques have been
introduced,
but they deal with only {\em pre-defined}
inductive heap predicates, whose definitions and semantics are given in
advance
\cite{BerdineCO04,BerdineCO05,CookHOPW11,PerezR13,PerezR11,BozgaIP10,PiskacWZ13,PiskacWZ14}. Since these techniques are designated to only certain classes of
pre-defined predicates, they are not suitable for handling general
inductive heap predicates.

Iosif et al.~\cite{IosifRS13,IosifRV14} and Enea et al.~\cite{EneaLSV14}
aim to prove entailments in more general SL fragments by translating SL
assertions into tree automata. However, these approaches still have certain
restrictions on inductive heap predicates, such as the predicates must have
the {\em bounded tree width} property, or they are variants of linked list
structures.
Proof techniques proposed by Nguyen et al.
\cite{NguyenDQC07,NguyenC08,ChinDNQ12}
and by Madhusudan et al.~\cite{QiuGSM13} can prove SL
entailments with {\em general} inductive heap predicates. Nonetheless,
these techniques are semi-automated since users are required to provide
supplementing lemmas to assist in handling those predicates.
In~\cite{EneaSW15}, Enea et al. develop a mechanism to automatically
synthesize these supporting lemmas, but solely limited to certain kinds of
lemmas, i.e., {\em composition lemmas}, {\em completion lemmas} and {\em
stronger lemmas\/}.

Cyclic proof ~\cite{BrotherstonDP11,BrotherstonGP12} and induction proof
in~\cite{ChuJT15} are most closely related to our approach. We recall the
aforementioned comments that cyclic proof requires soundness condition to
be checked globally on proof trees, whereas proof technique
in~\cite{ChuJT15} restricts that induction hypotheses collected from one
path of proof tree cannot be used to prove entailments in other paths. Our
work differs from them as we not only allow soundness condition to be
checked locally at inference rule level, but also support mutual induction
where entailments from different proof paths can be used as hypotheses to
prove each other.

{\bf Contribution.}
Our contributions in this work are summarized as follows:

\begin{enumerate}
  \setlength{\itemsep}{1pt}\setlength{\parskip}{3pt}

\item[--] We define a well-founded relation on SL models and use it to
  construct a novel mutual induction principle for proving SL entailments.

\item[--] We develop a deductive system for proving SL
  entailments based on the proposed mutual induction principle, and prove
  soundness of the proof system.

\item[--] We implement a prototype prover, named \textsf{Songbird}, and
  experiment on it with benchmarks of handcrafted entailments as well as
  entailments collected from \textsf{SL-COMP}, an SL competition. Our prover
  is available for both online use and download at: \sburl.

\end{enumerate}


\section{Motivating Example}
\label{sec:Motivation}

We consider the procedure $\code{traverse}$ in Fig. \ref{fig:ProgramRandomJump},
which traverses a linked list in an unusual way, by randomly jumping either one
or two steps at a time. In order to verify memory safety of this program,
automated verification tools such as ~\cite{CalcagnoDOY09,LeGQC14} will first
formulate the shape of the computer memory manipulated by $\code{traverse}$.
Suppose the initially discovered shape is represented by an {\em inductive heap
  predicate} $\hformp{\predTmp}{x}$ in SL, defined as:

\begin{wrapfigure}[11]{r}{0.48\textwidth}
  \def\indent{\hspace*{2em}}
  \centering
  \begin{tabular}{l}
    $\code{struct~\,node~\,\{\,struct~\,node~\, \cpointer{next};\}}$ \\
    \\[-0.8em]
    $\code{void~\,traverse\,(\,struct~\,node~\,\cpointer{x}\,)\,\{}$ \\
    $\code{\indent if ~\, (\,x \shorteqeq NULL\,) ~\, return;}$ \\
    $\code{\indent bool ~\, jump \,\shorteq\, random();}$ \\
    $\code{\indent if ~\, (jump ~\,\&\&~\, x{\rightarrow}next \,\shortneq\, NULL)}$ \\
    $\code{\indent\indent
    traverse(x{\rightarrow}next{\rightarrow}next);}$ \\
    $\code{\indent else ~\, traverse(x{\rightarrow}next); \,\}}$
  \end{tabular}
  \caption{A linked-list traversal algorithm with random jump}
  \label{fig:ProgramRandomJump}
\end{wrapfigure}

\begin{center}
\begin{tabular}{c}
  $\hformp{\predTmp} {x}
  \,~\triangleq~\,
  \predEmp
  ~\mtor~
  \exists u. (\hformn{x}{u} * \hformp{\predTmp}{u})
  ~\mtor~
  \exists u, v. (\hformn{x}{u} * \hformn{u}{v} * \hformp{\predTmp}{v})$
\end{tabular}
\end{center}

Intuitively, $\hformp{\predTmp}{x}$ covers three possible cases of the shape,
which can be an empty memory $\predEmp$ (when $\code{x \shorteqeq NULL}$), or be
recursively expanded by a single data structure $\hformn{x}{u}$ (when
$\code{traverse}$ jumps one step), or be recursively expanded by two
structures $\hformn{x}{u}$ and $\hformn{u}{v}$ (when $\code{traverse}$ jumps two
steps). Note that $\hformn{x}{u}$ and $\hformn{u}{v}$ are SL predicates modeling
the data structure $\code{node}$. Details about the SL syntax will be explained in
Section \ref{sec:Background}.

Since the derived shape is anomalous, the verifiers or users may want to
examine if it is actually a linked \underline{{\bf l}}ist
\underline{{\bf s}}egment, modeled by the following predicate:

\begin{center}
\begin{tabular}{c}
$\hformp{\predLs}{x,y}
\,~{\triangleq}~\, (\predEmp \wedge x{=}y) \mtor \exists w. (\hformn{x}{w} *
\hformp{\predLs}{w,y})$
\end{tabular}
\end{center}

This can be done by checking the validity of the following entailment:

\begin{center}
\begin{tabular}{c}
  $E ~\triangleq ~\hformp{\predTmp}{x} \entails \exists y.\,\hformp{\predLs}{x,y}$
\end{tabular}
\end{center}

In the semantics of SL, the entailment $E$ is said to be {\em valid}, if all
memory models satisfying $\hformp{\predTmp}{x}$ also satisfy $\exists
y.\,\hformp{\predLs}{x,y}$. To prove it by induction, $E$ is firstly recorded as
an induction hypothesis (IH), then the predicate $\hformp{\predTmp}{x}$ is
analyzed in each case of its definition, via a method called unfolding, to
derive new entailments $E_1,E_2,E_3$ as follows.

\begin{center}
\begin{tabular}{c}
  $E_1 ~\triangleq~ \predEmp \entails
    \exists y.\,\hformp{\predLs}{x,y}$ \qquad
  $E_2 ~\triangleq~ \hformn{x}{u} \,{*}\, \hformp{\predTmp}{u}
    \entails \exists y.\,\hformp{\predLs}{x,y}$\\[5pt]
  $E_3 ~\triangleq~ \hformn{x}{u} * \hformn{u}{v} * \hformp{\predTmp}{v}
    \entails \exists y.\,\hformp{\predLs}{x,y}$
\end{tabular}
\end{center}

The entailment $E_1$ can be easily proved by unfolding the predicate
$\hformp{\predLs}{x,y}$ in the right side
by its base case to obtain a valid entailment $\predEmp \entails \exists
y. (\predEmp \wedge x=y)$. On the contrary, the entailment $E_2$ can only be
proved by using the induction hypothesis $E$. Its (simplified) proof tree can be
depicted in Fig. \ref{fig:InductionProofTree}.

\begin{figure}[ht]
\begin{prooftree}
  \def\ScoreOverhang{0em}
  \def\defaultHypSeparation{\hskip 2em}
  \AxiomC{{\small{$ $}}}
  \def\extraVskip{3pt}
  \RightLabel{\scriptsize{{
    $(\rulePureEntail)$:
    Valid, proved by external provers, e.g. Z3.
  }}}
  \UnaryInfC{{\small{$
    true
    \entails
    \exists y,w.\,(u{=}w \wedge t{=}y)
  $}}}
  \def\extraVskip{3pt}
  \RightLabel{\scriptsize{{
    $(\ruleStarPred)$:
    Match and remove predicates $\hformp{\predLs}{u,t}$
    and $\hformp{\predLs}{w,y}$.
  }}}
  \UnaryInfC{{\small{$
    \hformp{\predLs}{u,t}
    \entails
    \exists y,w.\,(\hformp{\predLs}{w,y} \wedge u{=}w)
  $}}}
  \def\extraVskip{3pt}
  \RightLabel{\scriptsize{{
    $(\ruleStarData)$:
    Match and remove data nodes $\hformn{x}{u}$
    and $\hformn{x}{w}$.
  }}}
  \UnaryInfC{{\small{$
    \hformn{x}{u} \,{*}\, \hformp{\predLs}{u,t}
    \entails
    \exists y,w.\,(\hformn{x}{w} \,{*}\, \hformp{\predLs}{w,y})
  $}}}
  \def\extraVskip{3pt}
  \RightLabel{\scriptsize{{
    $(\rulePredIntroRight)$:
    Unfold $\hformp{\predLs}{x,y}$
    by its inductive case.
  }}}
  \UnaryInfC{{\small{$
    (E_4) ~ \hformn{x}{u} \,{*}\, \hformp{\predLs}{u,t}
    \entails
    \exists y.\,\hformp{\predLs}{x,y}
  $}}}
  \def\extraVskip{3pt}
  \RightLabel{
    \scriptsize{
    $(\ruleHypo)$:
    Apply IH $E$ with subst. $[u/x]$,
    rename $y$ to fresh $t$}.}
  \UnaryInfC{{\small{$
    (E_2) ~ \hformn{x}{u} \,{*}\, \hformp{\predTmp}{u}
    \entails
    \exists y.\,\hformp{\predLs}{x,y}
  $}}}
\end{prooftree}
\caption{Proof tree of $E_2$, using induction hypothesis $E$}
\label{fig:InductionProofTree}
\end{figure}

We can also prove $E_3$ by the same method,
i.e., applying the IH $E$, and its proof tree
is shown in Fig. \ref{fig:InductionProofTreeThree}.

\begin{figure}[h]
\begin{prooftree}

  \def\ScoreOverhang{0em}
  \def\defaultHypSeparation{\hskip 2em}
  \AxiomC{{\small{$ $}}}
  \def\extraVskip{3pt}
  \RightLabel{\scriptsize{{
    $(\rulePureEntail)$:
    Valid, proved by external prover, e.g. Z3.
  }}}
  \UnaryInfC{{\small{$
    true
    \entails
    \exists y,z,w.\,(u{=}z \wedge v{=}w \wedge t{=}y)
  $}}}
  \def\extraVskip{3pt}
  \RightLabel{\scriptsize{{
    $(\ruleStarPred)$:
    Remove predicates $\hformp{\predLs}{v,t}$
    and $\hformp{\predLs}{w,y}$.
  }}}
  \UnaryInfC{{\small{$
    \hformp{\predLs}{v,t}
    \entails
    \exists y,z,w.\,(\hformp{\predLs}{w,y} \wedge u{=}z \wedge v{=}w)
  $}}}
  \def\extraVskip{3pt}
  \RightLabel{\scriptsize{{
    $(\ruleStarData)$:
    Remove data nodes $\hformn{u}{v}$
    and $\hformn{z}{w}$.
  }}}
  \UnaryInfC{{\small{$
    \hformn{u}{v} * \hformp{\predLs}{v,t}
    \entails
    \exists y,z,w.\,(\hformn{z}{w} * \hformp{\predLs}{w,y} \wedge u{=}z)
  $}}}
  \def\extraVskip{3pt}
  \RightLabel{\scriptsize{{
    $(\rulePredIntroRight)$:
    Unfolding $\hformp{\predLs}{z,y}$
    by inductive case.
  }}}
  \UnaryInfC{{\small{$
    \hformn{u}{v} * \hformp{\predLs}{v,t}
    \entails
    \exists y,z.\,(\hformp{\predLs}{z,y} \wedge u{=}z)
  $}}}
  \def\extraVskip{3pt}
  \RightLabel{\scriptsize{{
    $(\ruleStarData)$:
    Remove data nodes $\hformn{x}{u}$
    and $\hformn{x}{z}$.
  }}}
  \UnaryInfC{{\small{$
    \hformn{x}{u} * \hformn{u}{v} * \hformp{\predLs}{v,t}
    \entails
    \exists y,z.\,(\hformn{x}{z} * \hformp{\predLs}{z,y})
  $}}}
  \def\extraVskip{3pt}
  \RightLabel{\scriptsize{{
    $(\rulePredIntroRight)$:
    Unfold  $\hformp{\predLs}{x,y}$
    by inductive  case.
  }}}
  \UnaryInfC{{\small{$
    \hformn{x}{u} * \hformn{u}{v} * \hformp{\predLs}{v,t}
    \entails
    \exists y.\,\hformp{\predLs}{x,y}
  $}}}
  \def\extraVskip{3pt}
  \RightLabel{\scriptsize{$(\ruleHypo)$:~\parbox{24em}{
    Apply IH $E$ with substitution $[v/x]$, \\
    and rename $y$ to $t$
  }}}
  \UnaryInfC{{\small{$
    (E_3) ~ \hformn{x}{u} * \hformn{u}{v} * \hformp{\predTmp}{v}
    \entails
    \exists y.\,\hformp{\predLs}{x,y}
    $}}}

\end{prooftree}
\caption{Ordinary proof tree of $E_3$, using induction hypothesis $E$}
\label{fig:InductionProofTreeThree}
\end{figure}

Using a different strategy, we observe that once $E_2$ is proved, entailments
derived during its proof, i.e., $E_2$ and $E_4$, can be used as hypotheses to
prove $E_3$. In this case, the new proof of $E_3$ is much simpler than the above
original induction proof, as demonstrated in Fig. \ref{fig:MutualProofTree}; the
proving process, therefore, is more efficient.

In the new proof tree, the entailment $E_4$ can be directly used as a hypothesis
to prove other entailments since it is already proven {\em valid} (see Fig.
\ref{fig:InductionProofTree}). However, when $E_2$ is applied to prove $E_3$,
thus prove $E$, it is not straightforward to conclude about $E$,
since the validity of $E_2$ is still {\em unknown}.
This is because the proof of $E_2$ in Fig.
\ref{fig:InductionProofTree} also uses $E$ as a hypothesis. Therefore,
$E$ and $E_2$ jointly form a {\em mutual induction} proof, in which
they can be used to prove each other. The theoretical principle of this proof
technique will be introduced in Section \ref{sec:MutualInduction}.

\begin{figure}
\begin{prooftree}
  \def\defaultHypSeparation{\hskip 2em}
  \def\ScoreOverhang{0em}
  \AxiomC{{\small{$ $}}}
  \def\extraVskip{3pt}
  \RightLabel{\scriptsize{{
    $(\rulePureEntail)$:
    Valid, proved by external provers, e.g., Z3.
  }}}
  \UnaryInfC{{\small{$
    true
    \entails
    \exists y.\,y=z
  $}}}
  \def\extraVskip{3pt}
  \RightLabel{\scriptsize{{
    $(\ruleStarPred)$:
    Remove predicates $\hformp{\predLs}{x,z}$
    and $\hformp{\predLs}{x,y}$.
  }}}
  \UnaryInfC{{\small{$
    \hformp{\predLs}{x,z}
    \entails
    \exists y.\,\hformp{\predLs}{x,y}
  $}}}
  \def\extraVskip{3pt}
  \RightLabel{\scriptsize{$(\ruleHypo)$:
	Apply $E_4$ with subst. $[r/t]$,
	and rename $y$ to $z$.
  }}
  \UnaryInfC{{\small{$
    \hformn{x}{u} \,{*}\, \hformp{\predLs}{u,r}
    \entails
    \exists y.\,\hformp{\predLs}{x,y}
  $}}}
  \def\extraVskip{3pt}
  \RightLabel{\scriptsize{$(\ruleHypo)$:~\parbox{24em}{
    Apply hypothesis $E_2$ with subst. $[u{/}x,v{/}u]$,\\
    and rename $y$ to $r$.
  }}}
  \UnaryInfC{{\small{$
    (E_3) ~ \hformn{x}{u} \,{*}\, \hformn{u}{v} \,{*}\, \hformp{\predTmp}{v}
    \,{\entails}\,
    \exists y.\,\hformp{\predLs}{x,y}
  $}}}
\end{prooftree}
\caption{New proof tree of $E_3$, using hypotheses $E_2$ and $E_4$}
\label{fig:MutualProofTree}
\end{figure}


\section{Theoretical background}
\label{sec:Background}

In this work, we consider the {\em symbolic-heap} fragment of separation logic
with arbitrary user-defined inductive heap predicates. We denote this logic
fragment as $\theorySLSH$. It is similar to those introduced in
\cite{IosifRS13,BrotherstonGKR16}, but extended with linear arithmetic
($\theoryLA$) to describe more expressive properties of the data structures,
such as size or sortedness. The syntax and semantics of the $\theorySLSH$
assertions and their entailments are introduced in this section.

\subsection{Symbolic-heap Separation Logic}

{\bf Syntax.} The syntax of our considered separation logic fragment
$\theorySLSH$ is described in Fig. \ref{fig:SyntaxMSSL}.  In
particular, the predicate $\predEmp$ represents an {\em empty}
memory. The {\em singleton} heap predicate
$\hformnt{\varSort}{x}{x_1{,}{...}{,}x_n}$ models an $n$-field single
data structure in memory where $x$ points-to; its data type is
represented by a unique {\em sort} $\varSort$\footnote{Note that for
  the simplicity of presenting the motivating example, we have removed
  the sort $\varSort$ from the SL singleton heap predicate denoting
  the data structure $\code{node}$.}  and values of its fields are
captured by $x_1{,}{...}{,}x_n$.  The {\em inductive} heap predicate
$\hformp{\predP}{x_1{,}{...}{,}x_n}$ models a recursively defined data
structure, which is formally defined in Definition
\ref{def:InductivePredicate}.  These three heap predicates, called
{\em spatial atoms}, compose the {\em spatial} assertions $\Sigma$ via
the separating conjunction operator $*$.  $\Pi$ denotes {\em pure}
assertions in linear arithmetic, which do not contain any spatial
atoms.

\setlength{\intextsep}{0em}
\begin{definition}[Inductive heap predicate]
\label{def:InductivePredicate}
A system of $k$ inductive heap predicates
$\predPi$ of arity $n_i$
and parameters $x^i_1,...,x^i_{n_i}$, with $i = 1, ..., k$,
are syntactically defined as follows:

\setlength{\intextsep}{0em}
\begin{center}
\begin{tabular}{l}
  $\Big\{\, \hformp{\predPi}{x^i_1, ..., x^i_{n_i}}$
  ~~$\triangleq$~~
  $\formF^i_1(x^i_1, ..., x^i_{n_i})
    \,\mtor \dots \mtor\,
    \formF^i_{m_i}(x^i_1, ..., x^i_{n_i}) \,\Big\}^k_{i\,=\,1}$
\end{tabular}
\end{center}

where $\formF^i_j(x^i_1,...,x^i_{n_i})$, with $1 \,{\leq}\, j
\,{\leq}\, m_i$, is a {\em definition case} of
$\hformp{\predPi}{x^i_1,...,x^i_{n_i}}$.
Moreover, $\formF^i_j$ is a {\em base case}
of $\predPi$, if it does not contain any predicate symbol which is (mutually)
recursively defined with $\predPi$; otherwise, it is an {\em inductive case}.
\end{definition}

\begin{figure}[h]
\begin{tabular}{c}
\begin{minipage}{0.45\textwidth}
\begin{adjustwidth}{0em}{}
\begin{tabular}{m{1em}m{1.5em}m{22em}l}

  \multicolumn{4}{l}{
    $c, x,\varSort,\predP$ resp. denote constants,
    variables, data sorts, and predicate symbols.
  }\\[2pt]

   $e$
   & $\defBNF/$
   & $c~|~x~|~{-}e~|~e_1{+}e_2~|~e_1{-}e_2$
   & Integer expressions \\[2pt]

   $a$
   & $\defBNF/$
   & $\valNil~|~x~$
   & Spatial expressions \\[2pt]

  $\Pi$
  & $\defBNF/$
  & $a_1 = a_2~|~a_1 \neq a_2~|~e_1 = e_2~|~e_1 \neq e_2~|$
  & Pure assertions \\[2pt]
  {} & & \multicolumn{2}{l}{
    $e_1 > e_2~|~e_1 \geq e_2~|~e_1 < e_2~|~e_1 \leq e_2~|$}\\
  {} & & \multicolumn{2}{l}{
    $\neg\Pi~|~\Pi_1 \,{\wedge}\, \Pi_2~|~\Pi_1 \,{\vee}\, \Pi_2~
    |~\Pi_1 \,{\Rightarrow}\, \Pi_2~|~
    \forall x .\Pi~|~\exists x .\Pi$} \\[2pt]

  $\Sigma$
  & $\defBNF/$
  & $\predEmp~|~\hformnt{\varSort}{x}{x_1{,}{...}{,}x_n}~
    |~ \hformp{\predP}{x_1{,}{...}{,}x_n}~
    |~ \Sigma_1 * \Sigma_2$
  & Spatial assertions\\[2pt]

  $F$
  & $\defBNF/$
  & $\Sigma ~|~ \Pi ~|~ \Sigma \wedge \Pi ~|~\exists x. F$
  & $\theorySLSH$ assertions
\end{tabular}
\end{adjustwidth}
\end{minipage}
\end{tabular}
\caption{Syntax of assertions in $\theorySLSH$}
\label{fig:SyntaxMSSL}
\end{figure}

\begin{definition}[Syntactic equivalence]
\label{def:SyntacticEquiv}
The syntactical equivalence relation
of two spatial assertions $\Sigma_1$ and $\Sigma_2$,
denoted as $\Sigma_1 \synequiv \Sigma_2$,
is recursively defined as follows:

\begin{tabular}{lll}
  -- $\predEmp \synequiv \predEmp$ \quad\quad
  &
  -- $\hformnt{\varSort}{u}{v_1{,}...{,}v_n} \synequiv
      \hformnt{\varSort}{u}{v_1{,}...{,}v_n}$ \quad\quad
  &
  -- $\hformp{\predP}{u_1{,}...{,}u_n} \synequiv
    \hformp{\predP}{u_1{,}...{,}u_n}$ \\[5pt]

  \multicolumn{3}{l}{
  -- If $\Sigma_1 \synequiv \Sigma'_1$ and $\Sigma_2 \synequiv \Sigma'_2$,
  then
  $\Sigma_1 * \Sigma_2 \synequiv \Sigma'_1 * \Sigma'_2$
  and
  $\Sigma_1 * \Sigma_2 \synequiv \Sigma'_2 * \Sigma'_1$
  }
\end{tabular}

\end{definition}


{\bf Semantics.}
The semantics of $\theorySLSH$ assertions are given in Fig. \ref{fig:SemanticsMSSL}.
Given a set $\setVar$ of variables,  $\setSort$ of sorts, $\setVal$ of values
and $\setLoc \subset \setVal$ of memory addresses,
a model of an assertion consists of:
\begin{itemize}
\item a {\em stack} model $s$, which is a function
  $s{:} ~ \setVar \rightarrow \setVal$. We write \evalForm{\Pi}{s}
to denote valuation of a pure
assertion $\Pi$ under the stack model $s$.
Note that the constant $\valNil \in \setVal \setminus \setLoc$
denotes dangling memory address.
\item a {\em heap} model $h$, which is a partial function
$h{:} ~ \setLoc \rightharpoonup_{\code{fin}}
(\setSort \rightarrow (\setVal ~ \code{list}))$.
$\funcDom{h}$ denotes domain of $h$, and
$\cardFunc{h}$ is cardinality of $\funcDom{h}$.
We follow Reynolds' semantics~\cite{Reynolds08} to
consider {\em finite} heap models, i.e., $\cardFunc{h} \,{<}\, \infty$.
$h \disjoins h'$ indicates that $h$ and $h'$ have disjoint domains,
i.e., $\funcDom{h} \,{\cap}\, \funcDom{h'} \,{=}\, \setempty$, and
$h \hunions h'$ is the union of two disjoint heap models $h, h'$,
i.e., $h \disjoins h'$.
\end{itemize}

\subsection{Entailments in $\theorySLSH$}

In this section, we formally define the $\theorySLSH$ entailments
and introduce a new concept of {\em model of entailments},
which will be used in the next section to construct
the well-founded relation in our induction principle.

\begin{definition}[Entailment]
An entailment between two assertions $F$ and $G$,
denoted as $F \entails G$,
is said to be {\em valid} (holds), iff
$s,h \satisfies F$ implies that $s,h \satisfies G$,
for all models $s,h$.
Formally,\\[0.5em]
\centerline{
$F \entails G$ is valid, iff~
$\forall s,h.(s,h \satisfies F \mtimply s,h \satisfies G)$
}
\end{definition}

Here, $F$ and $G$ are respectively called the {\em antecedent}
and the {\em consequent} of the entailment.
For simplicity, the entailment $F \entails G$
can be denoted by just $E$, i.e., $E \triangleq F \entails G$.

\begin{figure}[ht]
\begin{tabular}{c}
\begin{minipage}{\textwidth}
\begin{adjustwidth}{1em}{}
\begin{tabular}{m{10em}m{1em}l}

  $s,h \satisfies \Pi$
  & \hspace{-1.2em} iff
  & \hspace{-0.6em} $\evalForm{\Pi}{s}\,{=}\,\valTrue$
    and $\funcDom{h}\,{=}\,\setempty$
  \\ [0.2em]

  $s,h \satisfies \predEmp$
  & \hspace{-1.2em} iff
  & \hspace{-0.6em} $\funcDom{h}\,{=}\,\setempty$
  \\ [0.2em]

  $s,h \satisfies \hformnt{\varSort}{x}{x_1{,}{...}{,}x_n}$
  & \hspace{-1.2em} iff
  & $\evalVar{x}{s} {\in} \setLoc$
    and $\funcDom{h}\,{=}\,\{ \evalVar{x}{s} \}$
  \\[0.2em]

  {}
  &
  & \hspace*{4em} and $h(\evalVar{x}{s})\varSort\,{=}\,
    (\evalVar{x_1}{s}, ..., \evalVar{x_n}{s})$
  \\[0.2em]

  $s,h \satisfies \hformp{\predP}{x_1{,}{...}{,}x_n}$
  & \hspace{-1.2em} iff
  & $s,h \satisfies \hformp{R_i}{x_1{,}{...}{,}x_n}$,
    with $\hformp{R_i}{x_1{,}{...}{,}x_n}$ is one of
  \\[0.2em]
  {}
  &
  & \hspace*{4em} the definition cases of $\hformp{\predP}{x_1{,}{...}{,}x_n}$
  \\[0.2em]

  $s,h \satisfies \Sigma_1 * \Sigma_2$
  & \hspace{-1.2em} iff
  & there exist $h_1, h_2$ such that: $h_1 \disjoins h_2$,
  $h_1 \hunions h_2\,{=}\,h$
  \\[0.2em]
  {}
  &
  & \hspace*{4em} and $s,h_1 \satisfies \Sigma_1$ and
    $s,h_2 \satisfies \Sigma_2$
  \\[0.2em]

  $s,h \satisfies \Sigma \wedge \Pi$
  & \hspace{-1.2em} iff
  & $\evalForm{\Pi}{s}\,{=}\,\valTrue$ and $s,h \satisfies \Sigma$
  \\[0.2em]

  $s,h \satisfies \exists x .F$
  & \hspace{-1.2em} iff
  & $\exists v \,{\in}\, \setVal \, . \, [s|x{:}v],h \satisfies F$
  \\[0.2em]

\end{tabular}
\end{adjustwidth}
\end{minipage}
\end{tabular}
\caption{Semantics of assertions in $\theorySLSH$.
$[f|x{:}y]$ is a function like $f$ except that it returns $y$
for input $x$.}
\label{fig:SemanticsMSSL}
\end{figure}


\setlength{\intextsep}{0em}

\begin{definition}[Model and counter-model]
  \label{def:ModelCounterModel}
  Given an entailment $E \triangleq F \entails G$. An SL model $s,h$ is called a
  {\em model} of $E$, iff $s,h \satisfies F$ implies $s,h \satisfies G$. On the
  contrary, $s,h$ is called a {\em counter-model} of $E$, iff $s,h \satisfies F$
  and $s,h \nsatisfies G$.
\end{definition}

We denote $s,h \satisfies (F \entails G)$, or $s,h \satisfies E$, if $s,h$ is a
model of $E$. Similarly, we write $s,h \nsatisfies (F \entails G)$, or $s,h
\nsatisfies E$, if $s,h$ is a counter-model of $E$. Given a list of $n$
entailments $E_1,...,E_n$, we write $s,h \satisfies E_1,...,E_n$ if $s,h$ is a
model of {\em all} $E_1,...,E_n$, and $s,h \nsatisfies E_1,...,E_n$ if $s,h$ is
a counter-model of {\em some} $E_1,...,E_n$.


\section{Mutual induction proof for separation logic entailment using model order}
\label{sec:MutualInduction}

In this section, we first introduce the general schema of
{\em Noetherian induction, a.k.a. well-founded induction,}
and then apply it in proving SL entailments.

{\bf Noetherian induction~\cite{Bundy01}}.
Given a conjecture $\mathcal{P}(\alpha)$,
with $\alpha$ is a structure of type $\tau$,
the general schema of Noetherian induction on the structure $\alpha$ is

\setlength{\intextsep}{0.5em}

\begin{figure}[ht]
\begin{prooftree}
  \AxiomC{$
    \forall \alpha\,{:}\,\tau.~ (\forall \beta\,{:}\,\tau. ~
    \beta\,{\prec_{\tau}}\,\alpha \mtimply \mathcal{P}(\beta))
    \mtimply \mathcal{P}(\alpha))
    $}
  \UnaryInfC{$ \forall \alpha\,{:}\,\tau.~ \mathcal{P}(\alpha) $}
\end{prooftree}
\end{figure}

where $\prec_{\tau}$ is a well-founded relation on $\tau$, i.e., there is no
infinite descending chain, like $... \prec_{\tau} \alpha_n \prec_{\tau} ...
\prec_{\tau} \alpha_2 \prec_{\tau} \alpha_1$. Noetherian induction can be
applied for arbitrary type $\tau$, such as data structures or control flow.
However, success in proving a conjecture by induction is highly dependent on the
choice of the induction variable $\alpha$ and the well-founded relation
$\prec_{\tau}$.

{\bf Proving SL entailments using Noetherian induction}. We observe that an SL
entailment $E$ is said to be {\em valid} if $s,h \satisfies E$ for all model
$s,h$, given that the heap domain is finite, i.e., $\forall h. |h| \in \setNats$,
according to Reynolds' semantics~\cite{Reynolds08}. This inspires us to define a
well-founded relation among SL models, called {\em model order}, by comparing
size of their heap domains. To prove an SL entailment by Noetherian induction
based on this order, we will show that if all the smaller models satisfying the
entailment implies that the bigger model also satisfies the entailment, then
the entailment is satisfied by all models, thus it is valid. The model order and
induction principle are formally described as follows.

\begin{definition}[Model order]
  The {\em model order}, denoted by $\ltmodel$, of SL models is a
  binary relation defined as: $s_1,h_1 \ltmodel s_2,h_2$, if $|h_1| < |h_2|$.
\end{definition}

\begin{theorem}[Well-founded relation]
  The model order $\ltmodel$ of SL models is a well-founded relation.
\end{theorem}

\begin{proof}
  By contradiction, suppose that $\ltmodel$ were not well-founded, then there
  would exist an infinite descending chain: $... \ltmodel s_n,h_n \ltmodel ...
  \ltmodel s_1,h_1$. It follows that there would exist an
  infinite descending chain: $... < |h_n| < ... < |h_1|$. This is
  impossible since domain size of heap model is finite, i.e., $|h_1|, ...,
  |h_n|, ... \in \setNats$. \qedhere
\end{proof}

\begin{theorem}[Induction principle]
\label{thm:InductionPrinciple}
An entailment $E$ is valid, if for all model $s,h$,
the following holds:
$(\forall s',h'.~ s',h' \ltmodel s,h \mtimply \satentail{s',h'}{E})
\mtimply \satentail{s,h}{E}$. Formally:
\begin{prooftree}
\AxiomC{$
  \forall s,h.~ (\forall s',h'.~
  s',h' \ltmodel s,h \mtimply \satentail{s',h'}{E})
  \mtimply \satentail{s,h}{E}
$}
\UnaryInfC{$
  \forall s,h.~ \satentail{s,h}{E}
$}
\end{prooftree}
\end{theorem}

Since our induction principle is constructed on the SL model order, an induction
hypothesis can be used in the proof of any entailment whenever the decreasing
condition on model order is satisfied. This flexibility allows us to extend the
aforementioned principle to support {\em mutual induction}, in which multiple
entailments can participate in an induction proof, and each of them can be used
as a hypothesis to prove the other. In the following, we will introduce our {\em
  mutual induction principle}. Note that the induction principle in Theorem
\ref{thm:InductionPrinciple} is an instance of this principle, when only one
entailment takes part in the induction proof.

\begin{theorem}[Mutual induction principle]
\label{thm:MutualInduction}
Given $n$ entailments $E_1,...,E_n$. All of them are valid, if for all model
$s,h$, the following holds: $(\forall s',h'.~ s',h' \ltmodel s,h \mtimply
\satentail{s',h'}{E_1,...,E_n}) \mtimply \satentail{s,h}{E_1,...,E_n}$.
Formally:

\begin{prooftree}
\AxiomC{$
  \forall s,h.~ (\forall s',h'.~
  s',h' \ltmodel s,h \mtimply \satentail{s',h'}{E_1,...,E_n})
  \mtimply \satentail{s,h}{E_1,...,E_n}
  $}
\UnaryInfC{$
  \forall s,h.~ \satentail{s,h}{E_1,...,E_n}
  $}
\end{prooftree}
\end{theorem}

\begin{proof}

  By contradiction, assume that some of $E_1,...,E_n$ were invalid. Then, there
  would exist some counter-models $s,h$ such that $s,h \nsatisfies E_1,...,E_n$.
  Since $\ltmodel$ is a well-founded relation, there would exist the {\em least}
  counter-model $s_1,h_1$ such that $s_1,h_1 \nsatisfies E_1,...,E_n$,
  and, $s'_1,h'_1 \satisfies E_1,...,E_n$ for all $s'_1, h'_1
  \ltmodel s_1, h_1$. Following the theorem's hypothesis $\forall s,h.~ (\forall
  s',h'.~ s',h' \,{\ltmodel}\, s,h \mtimply \satentail{s',h'}{E_1,...,E_n})
  \mtimply \satentail{s,h}{E_1,...,E_n}$, we have $s_1,h_1 \satisfies
  E_1,...,E_n$. This contradicts with the assumption that $s_1,h_1$ is a
  counter-model. \qed

\end{proof}


\section{The proof system}
\label{sec:Implementation}

In this section, we introduce a sequent-based deductive system, which comprises a
set of inference rules depicted in Fig.~\ref{fig:LogicalRule} (logical rules)
and Fig.~\ref{fig:InductionRule} (induction rules), and a proof search procedure
in Fig.~\ref{fig:ProofSearch}. Each inference rule has zero or more premises, a
conclusion and possibly a side condition. A premise or a conclusion is described
in the same form of $\varHypo,~ \varTrace,~ F_1 \entails F_2$, where (i) $F_1
\entails F_2$ is an entailment, (ii) $\varHypo$ is a set of entailments with
validity status, which are recorded during proof search and can be used as
hypotheses to prove $F_1 \entails F_2$, and (iii) $\varTrace$ is a proof trace
capturing a chronological list of inference rules applied by the proof search
procedure to reach $F_1 \entails F_2$.

In addition, the entailment in the conclusion of a rule is called the {\em goal
  entailment}. Rules with zero (empty) premise is called {\em axiom rules}. A
proof trace $\varTrace$ containing $n$ rules $\ruleR_1, \ldots, \ruleR_n$, with
$n \,{\geq}\, 0$, is represented by $[(\ruleR_1), \ldots, (\ruleR_n)]$, where
the head $(\ruleR_1)$ of $\varTrace$ is the latest rule used by the proof search
procedure. In addition, some operations over proof traces are (i) insertion:
$(\ruleR)\,{\inserttrace}\,\varTrace$, (ii) membership checking: $(\ruleR)
\membertrace \varTrace$, and (iii) concatenation: $\varTrace_1 \concattrace
\varTrace_2$.

\setlength{\intextsep}{0.8em}
\begin{figure}[ht]
\begin{small}
\begin{minipage}{\textwidth}

  \begin{tabular}{ll}

    \minipageFalseLeftOne{0.29\textwidth}
    &
    \minipageFalseLeftTwo{0.48\textwidth}
    \\[1.5em]

    \minipagePureEntail{0.3\textwidth}
    &
    \minipageEmpLeftOne{0.41\textwidth}
    \\[1.5em]

    \minipageEqualLeft{0.3\textwidth}
    &
    \minipageEmpRightOne{0.47\textwidth}
    \\[1.8em]

    \minipageEqualRight{0.3\textwidth}
    &
    \hspace{-1.3em}
    \minipageStarData{0.65\textwidth}
    \\[1.8em]

    \minipageExistsLeft{0.35\textwidth}
    &
    \minipageStarPred{0.595\textwidth}
    \\[1.5em]

    \minipageExistsRight{0.2\textwidth}
    &
    \hspace{-1.8em}
    \minipagePredIntroRight{0.54\textwidth}
    \\[1.8em]
  \end{tabular}

\caption{Logical rules.
Note that for a rule $\ruleR$ with trace $\varTrace$ in its conclusion,
the trace in its premise is $\varTrace' \triangleq (\ruleR) \inserttrace \varTrace$.}

\label{fig:LogicalRule}
\end{minipage}
\end{small}
\end{figure}

\titlespacing*{\subsection}{0pt}{1ex plus .1ex}{0.5ex plus .1ex}

\subsection{Logical rules}

Logical rules in Fig.~\ref{fig:LogicalRule} deal with the logical structure of SL
entailments. For brevity, in these rules, we write the {\em complete\/}
symbolic-heap assertion $\exists \vec{x}. (\Sigma \wedge \Pi)$ as a {\em
  standalone\/} $F$. We define the {\em conjoined\/} assertion $F * \Sigma'
\triangleq \Sigma * \Sigma' \wedge \Pi$ and $F \wedge \Pi' \triangleq \Sigma \wedge
\Pi \wedge \Pi'$, given that existential quantifiers does not occur in the
outermost scope of $F$, i.e., $F \triangleq \Sigma \wedge \Pi$. The notation
$\vec{u}{=}\vec{v}$ means $(u_1{=}v_1) \,{\wedge}\, \ldots \,{\wedge}\,
(u_n{=}v_n)$, given that $\vec{u}{=}u_1{,} \ldots {,}u_n$ and $\vec{v}{=}v_1{,}
\ldots {,}v_n$ are two lists containing the same number of variables. We also
write $\vec{x} \disjoins \vec{y}$ to denote $\vec{x}$ and $\vec{y}$ are
disjoint, i.e., $\nexists u. (u \in \vec{x} \mtand u \in \vec{y})$, and use
\freevars{F} to denote the list of all free variables of an assertion $F$.
Moreover, $F[e/x]$ is a formula obtained from $F$ by
substituting the expression $e$ for all occurrences of the free variable
$x$ in $F$.

The set of logical rules are explained in details as follows:

\setlength{\intextsep}{0.0em}

\begin{enumerate}

  \setlength{\itemsep}{3pt}\setlength{\parskip}{0.5pt}

\item[--] {\bf Axiom rules.} The rule $\rulePureEntail$ proves a pure entailment
  $\Pi_1 \entails \Pi_2$ by invoking off-the-shelf provers such as
  Z3~\cite{MouraB08} to check the pure implication $\Pi_1 \,{\Rightarrow}\,
  \Pi_2$ in its side condition. The two rules $\ruleFalseLeftOne$ and
  $\ruleFalseLeftTwo$ decide an entailment {\em vacuously\/} valid if its
  antecedent is unsatisfiable, i.e., the antecedent contains a contradiction
  $(u{\neq}u)$ or overlaid data nodes $(\hformnt{\varSort_1}{u}{\vec{v}} *
  \hformnt{\varSort_2}{u}{\vec{w}})$.

\item[--] {\bf Normalization rules.} These rules simplify their goal
  entailments by either eliminating existentially quantified variables
  ($\ruleExistsLeft, \ruleExistsRight$), or removing equalities
  ($\ruleEqualLeft, \ruleEqualRight$) or empty heap predicates ($\ruleEmpLeft,
  \ruleEmpRight$) from antecedents (left side) or consequents (right side) of
  the entailments.

\item[--] {\bf Frame rules.} The two rules $\ruleStarData$ and $\ruleStarPred$
  applies the {\em frame property\/} of SL~\cite{Reynolds08} to remove {\em
    identical\/} spatial atoms from two sides of entailments. Note that the
  identical condition is guaranteed by adding equality constraints of these
  spatial atoms' arguments into consequents of the derived entailments.

\item[--] {\bf Unfolding rules.} The rule $\rulePredIntroRight$ derives a new
  entailment by unfolding a heap predicate in the goal entailment's consequent
  by its inductive definition. Note that unfolding a heap predicate in the
  entailment's antecedent will be performed by the induction rule
  $\ruleInduction$, as discussed in the next section.

\end{enumerate}

\setlength{\intextsep}{0.3em}

\subsection{Induction rules}

\setlength{\intextsep}{0em}
\begin{figure}[H]
\begin{small}

\begin{tabular}{l}
\begin{minipage}{0.6\textwidth}
\begin{prooftree}
  \def\defaultHypSeparation{\hskip 1em}
  \def\ScoreOverhang{0em}
  \AxiomC{$
    \varHypo \cup \{(H,\statusUnknown)\},\,
    \varTracePrim,\,
    F_1 * \hformp{\formF^{\predP}_1}{\vec{u}}
    \entails
    F_2
  $}
  \AxiomC{$\dots$}
  \AxiomC{$
    \varHypo \cup \{(H,\statusUnknown)\},\,
    \varTracePrim,\,
    F_1 * \hformp{\formF^{\predP}_m}{\vec{u}}
    \entails
    F_2
  $}
  \def\extraVskip{3pt}
  \LeftLabel{\rulename{\ruleInduction}}
  \RightLabel{$\dagger_{(\ruleInduction)}$}
  \TrinaryInfC{$
    \varHypo \sepAnte\, \varTrace \sepAnte\,
    F_1 * \hformp{\predP}{\vec{u}}
    \entails
    F_2
  $}
\end{prooftree}
\end{minipage}
\\
\begin{minipage}{\textwidth}
\begin{adjustwidth}{1.3em}{}
\begin{small}
Given
$H \triangleq F_1 * \hformp{\predP}{\vec{u}} \entails F_2$,
$\varTracePrim = (\ruleInduction) \inserttrace \varTrace$,
and
$\dagger_{(\ruleInduction)}$:~
$\hformp{\predP}{\vec{u}} \triangleq
\hformp{\formF^{\predP}_1}{\vec{u}} \mtor
 \ldots
 \mtor \hformp{\formF^{\predP}_m}{\vec{u}}$
\end{small}
\end{adjustwidth}
\end{minipage}
\end{tabular}
\\[0.8em]

\begin{tabular}{l}
\begin{minipage}{\textwidth}
\raggedleft{}
\begin{prooftree}
  \def\defaultHypSeparation{\hskip 2em}
  \def\ScoreOverhang{0em}
  \AxiomC{$
    \varHypo \cup \{(H, status)\},~
    (\ruleHypo) \inserttrace \varTrace,~
    F_4\theta \,{*}\, \Sigma' \,{\wedge}\, \Pi_1
    \entails
    F_2
  $}
  \def\extraVskip{3pt}
  \LeftLabel{\rulename{\ruleHypo}}
  \RightLabel{~\parbox{15em}{\rulesidecondright{
    \exists{} \theta{,} \Sigma'. (\Sigma_1 {\synequiv} \Sigma_3\theta{*} \Sigma'
    \mtand{} \Pi_1 {\Rightarrow} \Pi_3\theta)},\\
    \rulesidecondright{\dagger_{(\ruleHypo)}}
  }}
  \UnaryInfC{$
    \varHypo \,{\cup}\, \{(H \,{\triangleq}\, \Sigma_3 {\wedge} \Pi_3 {\entails} F_4, status)\},\,
    \varTrace,\,
    \Sigma_1 \,{\wedge}\, \Pi_1
    \,{\entails}\,
    F_2
  $}
\end{prooftree}
\end{minipage}
\\
\begin{minipage}{\textwidth}
\begin{adjustwidth}{1em}{}
\begin{small}
\begin{tabular}{ll}
with $\dagger_{(\ruleHypo)}$:
$(status {=} \statusValid)$
&
$\mtor
\exists \varSort, u, \vec{v}, \Sigma''.
(\Sigma' \synequiv \hformntShort{\varSort}{u}{\vec{v}} * \Sigma'')$\\
{}&
$\mtor
\exists \varTrace_1, \varTrace_2.(
\varTrace \,{=}\, \varTrace_1 {\concattrace} [(\ruleStarData)] {\concattrace}
\varTrace_2
\mtand (\ruleInduction) \,{\nmembertrace}\, \varTrace_1
\mtand (\ruleInduction) \,{\membertrace}\, \varTrace_2)$.\\
\end{tabular}
\end{small}
\end{adjustwidth}
\end{minipage}
\end{tabular}

\end{small}
\caption{Induction rules}
\label{fig:InductionRule}
\end{figure}

Fig.~\ref{fig:InductionRule} presents inference rules implementing
our mutual induction principle.
The {\em induction\/} rule $\ruleInduction$ firstly records its goal
entailment as an induction hypothesis $H$, and unfolds an inductive heap
predicate in the antecedent of $H$ to derive new entailments. When $H$ is inserted
into the hypothesis vault $\varHypo$, its status is initially assigned to
$\statusUnknown$ ({\em unknown\/}), indicating that its validity is not known at
the moment. Later, the status of $H$ will be updated to $\statusValid$ ({\em
  valid\/}) once the proof search procedure is able to prove it valid.
Generally, given an entailment $E$ and its proof tree $\mathcal{T}$, the proof
search procedure concludes that $E$ is valid if (i) every leaf of $\mathcal{T}$
is empty via applications of axiom rules, and (ii) all hypotheses used by the
{\em apply hypothesis\/} rule $\ruleHypo$ must be derived in $\mathcal{T}$.

Rule $\ruleHypo$ is the key rule of our mutual induction principle, which
applies an appropriate hypothesis $H \triangleq \Sigma_3 \wedge \Pi_3 \entails
F_4$ in proving its goal entailment $E \triangleq \Sigma_1 \wedge \Pi_1 \entails
F_2$. The rule firstly unifies the antecedents of $H$ and $E$ by a substitution
$\theta$, i.e., there exists a spatial assertion $\Sigma'$ such that $\Sigma_1
\synequiv \Sigma_3\theta * \Sigma'$ and $\Pi_1 \Rightarrow \Pi_3\theta$. If such
$\theta$ and $\Sigma'$ exist, we can weaken the antecedent of $E$ as follows $
(\Sigma_1 \wedge \Pi_1) \entails (\Sigma_3\theta * \Sigma' \wedge \Pi_3\theta
\wedge \Pi_1) \entails (F_4\theta * \Sigma' \wedge \Pi_1) $. Note that we use
Reynolds's substitution law~\cite{Reynolds08} to obtain $\Sigma_3\theta \wedge
\Pi_3\theta \entails F_4\theta$ from the hypothesis $H$. The proof system then
derives the next goal entailment $F_4\theta * \Sigma' \wedge \Pi_1 \entails F_2$
as shown in the premise of rule $\ruleHypo$.

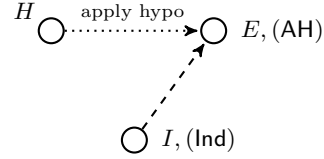
\begin{wrapfigure}[8]{r}{0.43\textwidth}
\begin{tikzpicture}[
  ->,>=stealth',shorten >=1pt,auto,node distance=2.4cm,thick,
  main node/.style={circle,draw,font=\sffamily},
  label/.style={draw=none},
  hidden node/.style={font=\sffamily}]

  \node[main node] (E) {};
  \node[label] [right =0.05cm of E] (L) {{\small $E, (\ruleHypo)$}};

 \node[main node] (H) [left =1.8cm of E] {};
  \node[label] [above left =-0.1cm and -0.05cm of H] (L) {{\small $H$}};

  \node[main node] (R) [below left=1.2cm and 0.8cm of E] {};
  \node[label] [right =0.05cm of R] (L) {{\small $I, (\ruleInduction)$}};

  \draw [rounded corners,dotted] (H) -- node{{\scriptsize apply hypo}} (E);
  \draw [rounded corners,dashed] (R) -- (E);
\end{tikzpicture}
\caption{Applying hypothesis}
\label{fig:ApplyingHypothesis}
\end{wrapfigure}

The side condition $\dagger_{(\ruleHypo)}$ of rule $\ruleHypo$ ensures the
decreasing condition of the mutual induction principle. In particular,
suppose that the proof search procedure applies a hypothesis $H$ in $\varHypo$
to prove an entailment $E$ via rule $\ruleHypo$. If the status of $H$ is
$\statusValid$, denoted by the first condition in $\dagger_{(\ruleHypo)}$, then
$H$ is already proved to be valid; thus it can be freely used to prove other
entailments. Otherwise, the status of $H$ is $\statusUnknown$, and $H$ may
participate in a (mutual) induction proof with an entailment $I$ in the proof
path of $E$, as depicted in Fig.~\ref{fig:ApplyingHypothesis}. Note that the
entailment $I$ has been recorded earlier as an induction hypothesis by an
application of the induction rule $\ruleInduction$.

In the latter case, the induction principle requires the decrease of model size
when applying the hypothesis $H$ to prove  entailment $I$. We then show
that this decreasing condition holds if one of the following conditions of
$\dagger_{(\ruleHypo)}$ is satisfied.

\setlength{\intextsep}{0em}

\begin{enumerate}

\setlength{\itemsep}{2pt}\setlength{\parskip}{0pt}

\item[(i)] $\exists \varSort, u, \vec{v}, \Sigma''. (\Sigma' {\synequiv}
  \hformntShort{\varSort}{u}{\vec{v}} {*} \Sigma'')$ indicates that the
  left-over heap part $\Sigma'$ after unifying antecedent of $H$ into that
  of $E$ contains at least one singleton heap predicate, or

\item[(ii)] $\exists \varTrace_1, \varTrace_2.( \varTrace {=} \varTrace_1
  {\concattrace} [(\ruleStarData)] {\concattrace} \varTrace_2 \mtand
  (\ruleInduction) {\nmembertrace} \varTrace_1 \mtand (\ruleInduction)
  {\membertrace} \varTrace_2)$ requires that there is a removal step of a
  singleton heap predicate by the rule $\ruleStarData$ applied
  between this hypothesis application $\ruleHypo$ and the most recent
  induction step $\ruleInduction$.

\end{enumerate}

Consider an arbitrary model $s,h$ satisfying $I$. During the derivation path from $I$ to
$E$, the model $s,h$ is transformed into a corresponding model $s_e,h_e$ of $E$.
We always have $|h_e| \leq |h|$ as the applications of logical rules and rule
$\ruleInduction$ never increase heap model size of entailments.
Moreover,
when applying $H$ to prove $E$, the model $s', h'$ of $H$, which
corresponds to $s_e, h_e$ of $E$,
satisfies $|h'| \leq |h_e|$, due to the unification step in
rule $\ruleHypo$. We consider two following cases. If condition
(i) is satisfied, then heap model size of the left-over part $\Sigma'$ is at
least 1 since $\Sigma'$ contains a singleton heap predicate. As a result, $|h'|
< |h_e|$ and it follows that $|h'| < |h|$. If condition (ii) is satisfied, then
$|h_e| < |h|$ since there is a singleton heap predicate, whose size of heap
model is 1, is removed when deriving $I$ to $E$. This implies that $|h'| < |h|$.
In summary, we obtain that $|h'| < |h|$ for both cases; thus, $s',h' \ltmodel
s,h$. This concludes our explanation about the rule $\ruleHypo$.

\vspace{1em}

\begin{figure}[H]
\begin{algorithm}[H]
\begin{small}
\caption*{{\bf Procedure} \psCall{\procProve}{\varHypo,\varTrace,F \entails{} G}}
\psKey{Input:}
  $\varHypo, F \entails G$ and  $\varTrace$
  are respectively a set of hypotheses, a goal entailment and its corresponding proof trace.\\
\psKey{Output:} Validity result ({$\valValid$} or {$\valUnknown$}),
a set of derived entailments with their validity statuses,
and a set of hypotheses used in proof of $F \entails G$.
\label{proc:ProveEntailment}
\begin{algorithmic}[1]
  \def\indent{\hspace{2em}}
  \State
  \psAssign{\varRuleSelected}
    {\setenum{\varRule_{inst}~|~
     \varRule_{inst} = \psCall{\proc{Unify}}{\varRule, (\varHypo, \varTrace, F \entails{} G)}
      ~\mtand~\varRule\in\varRuleSet}}
  \label{line:FindRule}

  \If{$\ruleRs = \setempty$}
    \psReturn{\valUnknown, \setempty, \setempty}
    \Comment{no rule is selected}
  \EndIf{}
  \label{line:NoRule}

  \For{\psKey{each} $\varRule_{inst}$ \psKey{in} $\ruleRs$}
    \If{\psCall{\proc{GetName}}{\varRule_{inst}} $\in$ $\{\rulePureEntail, \ruleFalseLeftOne, \ruleFalseLeftTwo\}$ }
      \Comment{$\varRule$ is an axiom rule}
      \State\psReturn{\valValid, \setempty, \setempty}
    \EndIf{}
    \label{line:AxiomRule}

    \State\psAssign{\varHypo_{used}}{\setempty}
    \label{line:UsedHypoOne}

    \If{$\varRule_{inst} = \ruleHypo$ with hypothesis $E$}
      \psAssign{\varHypo_{used}}{\varHypo_{used} \cup{} \{E\}}
    \EndIf{}
    \label{line:UsedHypoTwo}

    \State\psAssign{\varHypo_{derived}}{\setempty}
    \label{line:DerivedEntailOne}

    \State \psAssign{(\varHypo_i,\varTrace_i,F_i \entails G_i)_{i=1, \ldots, n}}{
      \psCall{\proc{GetPremises}}{\varRule_{inst}}}
      \Comment{all premises of $\varRule_{inst}$}
      \label{line:PremiseBegin}

      \For{i = 1 \psKey{to} n}
        \State \psAssign{res,\varHypo_{derived},\varHypo'_{used}}{
          \psCall{\procProve}
          {\varHypo_i \oplus \varHypo_{derived},\varTrace_i,F_i \entails G_i}}
        \label{line:MutualInduction}
        \label{line:UsedHypoThree}
        \label{line:DerivedEntailTwo}
        \If{$res = {\valUnknown}$}
          \psReturn{\valUnknown, \setempty, \setempty}
        \label{line:OnePremiseFail}
        \EndIf{}
        \State{\psAssign{\varHypo_{used}} {\varHypo_{used} \cup{} \varHypo'_{used}}}
        \label{line:UsedHypoFour}
      \EndFor{}

        \If{$\varHypo_{used} \subseteq
             (\psCall{\proc{GetEntailments}}{\varHypo_{derived}} \cup \{ F \entails G\})$}
        \label{line:DecideStatusValidOne}
          \State{\psAssign{\varHypo_{derived}}{\varHypo_{derived} \oplus{} \{(F \entails{} G, \statusValid)\}}}
        \label{line:DecideStatusValidTwo}
        \label{line:DerivedEntailThree}
        \Else
          \,\psAssign{\varHypo_{derived}}{\varHypo_{derived} \oplus \{(F \entails G, \statusUnknown)\}}
        \label{line:DerivedEntailFour}
        \label{line:DecideStatusValidThree}
        \EndIf{}

      \State{\psReturn{\valValid, \varHypo_{derived}, \varHypo_{used}}}
      \label{line:AllPremisesValid}
      \Comment{all derived premises are proved}
  \EndFor{}
  \label{line:PremiseEnd}

  \State{\psReturn{\valUnknown, \setempty, \setempty}}
  \Comment{all rules fail to prove $F \entails G$}
\label{line:AllRuleFail}

\label{line:ProveEntailmentEnd}
\end{algorithmic}
\end{small}
\end{algorithm}
\caption{General proof search procedure,
in which $\varRuleSet$ is the set of inference rules given in
Fig.~\ref{fig:LogicalRule} and~\ref{fig:InductionRule}.}
\label{fig:ProofSearch}
\end{figure}

\titlespacing*{\subsection}{0pt}{1ex plus .1ex}{0.5ex plus .1ex}

\subsection{Proof search procedure}

Our proof search procedure $\procProve$ is designed in a self-recursive manner,
as presented in Fig.~\ref{fig:ProofSearch}. Its inputs consist of a set of hypotheses,
a proof trace, and an entailment, which are components of an inference
rule's conclusion. To prove a candidate entailment $F \entails G$, initially the
hypothesis set $\varHypo$ and the proof trace are assigned to empty ($\setempty$
and $[\,]$).

Firstly, the procedure $\procProve$ finds a set $\varRuleSelected$ of suitable
rules, whose conclusion can be unified with the goal entailment $F \entails G$,
among all inference rules in $\varRuleSet$ (line~\ref{line:FindRule}). If no
suitable rule is found, the procedure immediately returns $\valUnknown$,
indicating that it is unable to prove the entailment (line~\ref{line:NoRule}).
Otherwise, it subsequently processes each discovered rule $\varRule_{inst}$ in
$\varRuleSelected$ by either (i) returning $\valValid$ to announce a valid
result, if an axiom rule is selected (line \ref{line:AxiomRule}), or (ii)
recursively searching for proofs of the derived entailments in the premises of
$\varRule_{inst}$ (lines \ref{line:PremiseBegin}--\ref{line:PremiseEnd}). In the
latter case, the procedure returns $\valUnknown$ if one of the derived
entailments is not proved (line~\ref{line:OnePremiseFail}), or returns
$\valValid$ if all of them are proved (line~\ref{line:AllPremisesValid}).
Finally, it simply returns $\valUnknown$ when it cannot prove the goal
entailment with all selected rules (line~\ref{line:AllRuleFail}).

The procedure uses a local variable $\varHypo_{used}$ to store all hypotheses
used during the proof search. $\varHypo_{used}$ is updated when the rule
$\ruleHypo$ is applied (line~\ref{line:UsedHypoTwo}) or after the procedure
finishes proving a derived entailment (lines ~\ref{line:UsedHypoThree}
and~\ref{line:UsedHypoFour}). We also use another variable $\varHypo_{derived}$
to capture all generated entailments with their validity statuses.
The condition at line~\ref{line:DecideStatusValidOne} checks if all
hypotheses used to prove the entailment $F \entails G$ are only introduced
during the entailment's proof. If this condition is satisfied, then $F \entails
G$ is updated with a {\em valid status\/} $\statusValid$
(line~\ref{line:DecideStatusValidTwo}). Otherwise, the entailment may
participate in a (mutual) induction proof, thus its status is assigned to {\em
  unknown\/} $\statusUnknown$ (line \ref{line:DecideStatusValidThree}).

At line~\ref{line:MutualInduction}, the procedure uses not only the hypothesis
set $\varHypo_i$, introduced by the selected inference rule, but also the set
$\varHypo_{derived}$ containing entailments derived during proof search to prove
a new goal entailment $F_i \entails G_i$. This reflects our mutual induction
principle which allows derived entailments to be used as hypotheses in other
entailments' proofs. Note that the {\em union and update\/} operator $\oplus$
used in the algorithm will insert new entailments and their statuses into the
set of hypotheses, or update the existing entailments with their new statuses.
In addition, the auxiliary procedures used in our proof search procedure are
named in a self-explanatory manner. In particular, $\proc{Unify}$,
$\proc{GetName}$ and $\proc{GetPremises}$ respectively unifies an inference rule
with a goal entailment, or returns name and premises of an inference rule.
Finally, $\proc{GetEntailments}$ returns all entailments stored in the set of
derived entailments $\varHypo_{derived}$.

{\bf Soundness}. Soundness of our proof system is stated in Theorem
\ref{thm:Soundness}. Due to page constraint, we present the detailed proof in
the Appendix \ref{app:SoundnessInferenceRule}

\setlength{\intextsep}{0em}
\begin{theorem}[Soundness]
\label{thm:Soundness}
Given an entailment $E$, if the proof search procedure returns $\valValid$
when proving $E$, then $E$ is valid.
\end{theorem}


\section{Experiment}
\label{sec:Experiment}

We have implemented the proposed induction proof technique into a prototype
prover, named $\songbird$. The proof system and this paper's artifact are
available for both online use and download at {\sburl}.

\setlength{\intextsep}{0.6em}
\begin{figure}[H]
\begin{small}
\begin{tabular}{cc}
\begin{minipage}{0.6\textwidth}

\def\arraystretch{1.}
\setlength{\tabcolsep}{1.5pt}
\begin{tabular}{|l|c|c|c|c|c|}

\hline

\multicolumn{1}{|c|}{{\bf Category}} &
{\bf \slide} &
{\bf \spen} &
{\bf \sleek} &
{\bf \cyclist} &
{\bf $\songbird$}
\\

\hline

~\textsf{singly-ll} \hfill (64) & 12 & 3 & 48 & {\bf 63} & {\bf 63} \\

~\textsf{doubly-ll} \hfill  (37) & 14 & 0 & 17 & 24 & {\bf 26} \\

~\textsf{nested-ll} \hfill  (11) & 0 & {\bf 11} & 5 & 6 & {\bf 11} \\

~\textsf{skip-list} \hfill  (13) & 0 & {\bf 12}  & 4 & 5 & 7 \\

~\textsf{tree} \hfill  (26) & 12 & 1 & 14 & 18 & {\bf 22} \\

\hline
\hline

~Total \hfill  (151) & 38 & 27 & 88 & 116 & {\bf 129} \\

\hline

\end{tabular}
\end{minipage}

&

\begin{minipage}{0.4\textwidth}
\def\arraystretch{1.2}
\setlength{\tabcolsep}{1pt}
\begin{tabular}{|l|c|c|c|c|}

\hline

{} & \multicolumn{4}{c|}{{\bf $\songbird$}} \\

\cline{2-5}

&
$\cmarksb\,\xmarkother$ &
$\xmarksb\,\cmarkother$ &
$\cmarksb\,\cmarkother$ &
$\xmarksb\,\xmarkother$
\\

\hline

{\bf \cyclist} & 13 & 0 & 116 & 22 \\

\hline

{\bf \sleek} & 41 & 0 & 88 & 22 \\

\hline

{\bf \spen} & 109 & 7 & 20 & 15 \\

\hline

{\bf \slide} & 103 & 12 & 26 & 10 \\

\hline
\end{tabular}
\end{minipage}
\\
(a) & (b)

\end{tabular}
\end{small}
\caption{Overall evaluation on the benchmark
\textsf{slrd\_entl} of \textsf{SL-COMP}}
\label{fig:ExpSlcompAll}
\end{figure}

To evaluate our
technique, we compared our system against state-of-the-art SL provers, including
{\slide}~\cite{IosifRS13,IosifRV14}, {\spen}~\cite{EneaLSV14},
{\sleek}~\cite{ChinDNQ12} and {\cyclist}~\cite{BrotherstonDP11,BrotherstonGP12},
which had participated in the recent SL competition
\textsf{SL-COMP}~\cite{SighireanuC14}. We are however unable to make direct
comparison with the induction-based proof technique presented in \cite{ChuJT15}
as their prover was not publicly available. Our evaluation was performed on an
Ubuntu 14.04 machine with CPU Intel E5-2620 (2.4GHz) and RAM 64GB.

Firstly, we conduct the experiment on a set of {\em valid}
entailments\footnote{We exclude the set of invalid entailments because some
  evaluated proof techniques, such as~\cite{ChinDNQ12,BrotherstonDP11}, aim to
  only prove validity of entailments.}, collected from the benchmark
\textsf{slrd\_entl}\footnote{Available at
  \textsf{https://github.com/mihasighi/smtcomp14-sl/tree/master/bench}.} of
\textsf{SL-COMP}. These entailments contain {\em general} inductive heap predicates
denoting various data structures, such as singly linked lists
(\textsf{singly-ll}), doubly linked lists (\textsf{doubly-ll}), nested lists
(\textsf{nested-ll}), skip lists (\textsf{skip-list}) and trees (\textsf{tree}).
We then categorize problems in this benchmark based on their predicate types. In
Fig. \ref{fig:ExpSlcompAll}(a), we report the number of entailments successfully
proved by a prover in each category, with a timeout of 30 seconds for proving an
entailment. For each category, the total number of problems is put in
parentheses, and the maximum number of entailments that can be proved by the
list of provers are highlighted in bold. As can be seen, $\songbird$ can prove
more entailments than all the other tools. In particular, we are the best in
almost categories, except for \textsf{skip-list}. However, in this category, we
are behind only {\spen}, which has been specialized for skip lists
\cite{EneaLSV14}. Our technique might require more effective generalization to
handle the unproven \textsf{skip-list} examples.

In Fig. \ref{fig:ExpSlcompAll}(b), we make a detailed comparison among Songbird
and other provers. Specifically, the first column ($\cmarksb\,\xmarkother$)
shows the number of entailments that $\songbird$ can prove valid whereas the
others cannot. The second column ($\xmarksb\,\cmarkother$) reports the number of
entailments that can be proved by other tools, but not by $\songbird$. The last
two columns list the number of entailments that both Songbird and others can
($\cmarksb\,\cmarkother$) or cannot ($\xmarksb\,\xmarkother$) prove. We would
like to highlight that our prover efficiently proves {\em all} entailments
proved by {\cyclist} (resp. {\sleek}) in {\em approximately half the time},
i.e., 20.92 vs 46.40 seconds for 116 entailments, in comparison with {\cyclist}
(resp. 8.38 vs 15.50 seconds for 88 entailments, in comparison with {\sleek}).
In addition, there are 13 (resp. 41) entailments that can be proved by our tool,
but {\em not} by {\cyclist} (resp. {\sleek}). Furthermore, our $\songbird$
outperforms {\spen} and {\slide} by more than 65\% of the total entailments,
thanks to the proposed mutual induction proof technique.

Secondly, we would like to highlight the efficiency of {\em mutual induction} in
our proof technique via a comparison between $\songbird$ and its variant
$\songbirdSI$, which exploits only induction hypotheses found within a {\em
  single} proof path. This mimics the structural induction technique which
explores induction hypotheses in the same proof path. For this purpose, we
designed a new entailment benchmark, namely \textsf{slrd\_ind}, whose problems
are more complex than those in the \textsf{slrd\_entl} benchmark. For example,
our handcrafted benchmark\footnote{The full benchmark is available at \sburl.}
contains an entailment $\hformp{\predLsEven}{x,y} * \hformn{y}{z} *
\hformp{\predLsEven}{z,t} \entails \exists u.\,\hformp{\predLsEven}{x,u} *
\hformn{u}{t}$ with the predicate $\hformp{\predLsEven}{x,y}$ denoting list
segments with even length. This entailment was inspired by the entailment
$\hformp{\predLsEven}{x,y} * \hformp{\predLsEven}{y,z} \entails
\hformp{\predLsEven}{x,z}$ in the problem \textsf{11.tst.smt2} of
\textsf{slrd\_entl}, contributed by team {\cyclist}. Note that entailments in
our benchmark were constructed on the same set of linked list predicates
provided in \textsf{slrd\_entl}, comprised of regular singly linked lists
(\textsf{ll}), linked lists with even or odd length (\textsf{ll-even/odd}) and
linked list segments which are left- or right-recursively defined
(\textsf{ll-left/right}). We also use a new \textsf{ll2} list segment predicate
whose structure is similar to the predicate \textsf{tmp} in our motivating
example. In addition, problems in the \textsf{misc.} category involve all
aforementioned linked list predicates.

As shown in Fig. \ref{fig:ExpHandcraftedEntails}, $\songbirdSI$ is able to prove
nearly 70\% of the total entailments, which is slightly better than
{\cyclist}\footnote{We do not list other provers in Fig.
  \ref{fig:ExpHandcraftedEntails} as they cannot prove any problems in
  \textsf{slrd\_ind}.}, whereas $\songbird$, with full capability of mutual
induction, can prove the {\em whole} set of entailments. This result is
encouraging as it shows the usefulness and essentials of our mutual explicit
induction proof technique in proving SL entailments.

\setlength{\intextsep}{0.8em}

\begin{figure}[H]
\begin{center}
\def\arraystretch{1.1}
\begin{tabular}{|l|c|c|c|}

\hline

\multicolumn{1}{|c|}{{\bf Category}} &
{\bf \cyclist} &
{\bf $\songbird_\textbf{SI}$} &
{\bf $\songbird$}
\\

\hline

~ \textsf{ll/ll2} \hfill (24) & 18 & 22 & 24 \\

~ \textsf{ll-even/odd} \hfill (20) & 8 & 17 & 20 \\

~ \textsf{ll-left/right} \hfill (20) & 12 & 10 & 20 \\

~ \textsf{misc.} \hfill (32) & 17 & 16 & 32 \\

\hline

~ Total \hfill (96) & 55 & 65 & 96 \\

\hline

\end{tabular}
\end{center}
\caption{Comparison on \textsf{slrd\_ind} benchmark}
\label{fig:ExpHandcraftedEntails}
\end{figure}

\section{Conclusion}

We have proposed a novel induction technique and developed a proof system for
automatically proving entailments in a fragment of SL with general inductive
predicates. In essence, we show that induction can be performed on the size of
the heap models of SL entailments. The implication is that, during automatic
proof construction, the goal entailment and entailments derived in the entire
proof tree can be used as hypotheses to prove other derived entailments, and
vice versa. This novel proposal has opened up the feasibility of mutual
induction in automatic proof, leading to shorter proof trees being built.
In future, we would like to develop a verification system on top of the prover
{$\songbird$}, so that our {\em mutual explicit induction} technique can be
effectively used for automated verification of memory safety in imperative
programs.


{\bf Acknowledgement}. We would like to thank the anonymous reviewers for their
valuable and helpful feedback. The first author would like to thank Dr. James
Brotherston for the useful discussion about the cyclic proof. This work has been
supported by NUS Research Grant R-252-000-553-112. Ton Chanh and Wei-Ngan are
partially supported by MoE Tier-2 grant MOE2013-T2-2-146.

\bibliographystyle{splncs03}  
\bibliography{main}

\newpage
\appendix
\clearpage
\counterwithin{theorem}{section}  

\section{Soundness proof}
\label{app:SoundnessInferenceRule}
\counterwithin{theorem}{section}  
\counterwithin{proposition}{section}  

\subsection{Soundness of inference rules}

\setcounter{myequationNo}{0}

We prove soundness of these rules by showing that if entailments in their
premises are valid, and their side conditions are satisfied, then goal
entailments in their conclusions are also valid.

{\bf 1. Axiom rules $\ruleFalseLeftOne,\ruleFalseLeftTwo$ and $\rulePureEntail$:}

\begin{tabular}{ll}

  \minipageFalseLeftOne{0.3\textwidth}
  &
    \minipageFalseLeftTwo{0.55\textwidth}
  \\[0.8em]

  \minipagePureEntail{0.3\textwidth}
\end{tabular}

-- It is easy to verify that antecedents of goal entailments in the two rules
$\ruleFalseLeftOne,\ruleFalseLeftTwo$ are unsatisfiable, since they either
contain a contradiction ($u{\neq}u$ in the rule $\ruleFalseLeftOne$) or
contain two data nodes having the same memory address
($\hformntShort{\varSort_1}{u}{\vec{v}} *
\hformntShort{\varSort_2}{u}{\vec{w}}$ in the rule $\ruleFalseLeftTwo$).
Therefore, these entailments are evidently valid.

-- When the rule $\rulePureEntail$ is applied, pure provers will be invoked to check
the side condition: $\Pi_1 \Rightarrow \Pi_2$. If this side condition holds,
then clearly the entailment $\Pi_1 \entails \Pi_2$ is valid. \qedhere

{\bf 2. Rule $\ruleEqualLeft$ and $\ruleEqualRight$:}

\begin{tabular}{ll}

  \minipageEqualLeft{0.4\textwidth}
  &
    \minipageEqualRight{0.5\textwidth}
  \\[1.8em]
\end{tabular}

-- Soundness of the rule $\ruleEqualRight$ is evident since the condition
$u{=}u$ in the goal entailment $F_1 \entails \exists \vec{x}.(F_2 \wedge u{=}u)$
is a tautology.

-- For the rule $\ruleEqualLeft$, consider an arbitrary model $s,h$ of
antecedent of the goal entailment. Then $s,h \satisfies F_1 \wedge u{=}v$. It
follows that $\evalVar{u}{s} = \evalVar{v}{s}$, therefore $s,h \satisfies
F_1[u/v]$. Since the entailment $F_1[u/v] \entails F_2 [u/v]$ in the rule's premise
is valid, it implies $s,h \satisfies F_2[u/v]$. But this also means that $s,h
\satisfies F_2$, since $\evalVar{u}{s} = \evalVar{v}{s}$. Therefore, $F_1 \wedge
u{=}v \entails F_2$ is valid. \qedhere

{\bf 3. Rule $\ruleExistsLeft$ and $\ruleExistsRight$:}

\begin{tabular}{ll}

  \minipageExistsLeft{0.35\textwidth}
  &
  \minipageExistsRight{0.5\textwidth}
\end{tabular}

-- To prove correctness of the rule $\ruleExistsLeft$, we consider an arbitrary
model $s,h$ such that $s,h \satisfies \exists x. F_1$. By semantics of the
$\exists$ quantification, there is an integer value $v \in \setInt$ such that
$s', h \satisfies F_1$, with $s' = \modelExtOne{s}{x}{v}$. Then $s'', h
\satisfies F_1[u/x]$ with $s''$ is extended from $s'$ such that $s''(u) =
s'(x)$. Since $s' = \modelExtOne{s}{x}{v}$, it follows that $s'' =
\modelExtOne{s}{u}{v}$. On the other hand, given that entailment in the rule's
premise is valid, then $s'',h \satisfies F_2$. It implies that
$\modelExtOne{s}{u}{v} \satisfies F_2$. In addition $u \not\in \freevars{F_2}$.
Therefore $s,h \satisfies F_2$. Since $s,h$ is chosen arbitrarily, it follows
that the rule $\ruleExistsLeft$ is correct.

-- Correctness of $\ruleExistsRight$ is straight forward. Suppose that $s,h$ is
an arbitrary model such that $s,h \satisfies F_1$. Since entailment in the
rule's premise is valid, then $s,h \satisfies F_2[e/x]$. It follows that $s,h$
also satisfies $\exists x. F_2$, by simply choosing value $v$ of $x$ such that
$v = \evalForm{e}{s}$. Since $s,h$ is chosen arbitrarily, it follows that the
entailment in the rule's conclusion is valid. Therefore $\ruleExistsRight$ is
valid. \qedhere

{\bf 4. Rule $\ruleEmpLeft$ and $\ruleEmpRight$:}

\begin{tabular}{ll}

  \minipageEmpLeftOne{0.41\textwidth}
  &
    \minipageEmpRightOne{0.47\textwidth}
  \\[1.8em]
\end{tabular}

It is evident that two assertions $F_1 * \predEmp$ and $F_1$ in the rule
$\ruleEmpLeft$ are semantically equivalent. In addition, $F_2 * \predEmp$ and
$F_2$ in the rule $\ruleEmpRight$ are also semantically equivalent. It follows
that both the two rules $\ruleEmpLeft$ and $\ruleEmpRight$ are correct.
\qedhere

{\bf 5. Rule $\ruleStarData$ and $\ruleStarPred$:}

\begin{tabular}{l}

  \minipageStarData{0.65\textwidth}\\
  \minipageStarPred{0.595\textwidth}
  \\[1.5em]

\end{tabular}

In the following, we present soundness proof of the rule $\ruleStarData$.
Soundness of $\ruleStarPred$ can proved in a similar way.

Consider an arbitrary model $s,h$ such that $s,h \satisfies F_1 *
\hformntShort{\varSort}{u}{\vec{v}}$. Then, there exists $h_1 \disjoins h_2$
such that $h = h_1 \hunions h_2$, and $s,h_1 \satisfies F_1$, and $s,h_2
\satisfies \hformntShort{\varSort}{u}{\vec{v}}$. On one hand, entailment in the
rule's premise is valid, it follows that $s,h_1 \satisfies \exists \vec{x}. (F_2
\wedge u{=}t \wedge \vec{v}{=}\vec{w})$. By semantics of $\exists$
quantification, $s',h_1 \satisfies F_2 \wedge u{=}t \wedge \vec{v}{=}\vec{w}$,
with $s'$ is a model extended from $s$ with integer values of $\vec{x}$. On the
other hand, the rule's side condition gives $u \not\in \vec{x}$, and $\vec{v}
\disjoins \vec{x}$, and $s'$ is extend from $s$ with values of $\vec{x}$, and
$s,h_2 \satisfies \hformntShort{\varSort}{u}{\vec{v}}$, it follows that $s',h_2
\satisfies \hformntShort{\varSort}{u}{\vec{v}}$. Combining these two hands, with
the fact that $h_1 \disjoins h_2$ and $h = h_1 \hunions h_2$, the following
holds: $s',h \satisfies F_2 * \hformntShort{\varSort}{u}{\vec{v}} \wedge u{=}t
\wedge \vec{v}{=}\vec{w}$. By semantics of equality $(=)$, the following also
holds: $s',h \satisfies F_2 * \hformntShort{\varSort}{t}{\vec{w}} \wedge u{=}t
\wedge \vec{v}{=}\vec{w}$. By weakening this assertion via dropping the
condition $u{=}t \wedge \vec{v}{=}\vec{w}$, it is evident that $s',h \satisfies
F_2 * \hformntShort{\varSort}{t}{\vec{w}}$. Since $s'$ is extended from $s$ with
values of $\vec{x}$, it is evident that $s,h \satisfies \exists \vec{x}. (F_2 *
\hformntShort{\varSort}{t}{\vec{w}})$. Recall that $s,h$ is chosen arbitrarily,
this implies that the rule $\ruleStarData$ is sound. \qedhere

{\bf 6. Rule $\rulePredIntroRight$:}

\begin{tabular}{ll}
  \minipagePredIntroRight{0.54\textwidth}
\end{tabular}

Consider an arbitrary model $s,h$ such that $s,h \satisfies F_1$. Since
entailment in the rule's premise is valid, it follows that $s,h \satisfies
\exists \vec{x}. (F_2 * \hformp{\formF^{\predP}_i}{\vec{u}})$. In addition, the
rule's side condition that $\hformp{\formF^{\predP}_i}{\vec{u}}$ is one of the
definition cases of $\hformp{\predP}{\vec{u}}$ clearly implies that $s,h
\satisfies \exists \vec{x}. (F_2 * \hformp{\predP}{\vec{u}})$. Since $s,h$ is
chosen arbitrarily, it follows that entailment in the rule's conclusion is
valid. \qedhere

{\bf 7. Rule $\ruleInduction$:}

\begin{small}
\begin{tabular}{l}
\begin{minipage}{0.6\textwidth}
\begin{prooftree}
  \def\defaultHypSeparation{\hskip 1em}
  \def\ScoreOverhang{0em}
  \AxiomC{$
    \varHypo \cup \{(H,\statusUnknown)\},\,
    \varTracePrim,\,
    F_1 * \hformp{\formF^{\predP}_1}{\vec{u}}
    \entails
    F_2
  $}
  \AxiomC{$\dots$}
  \AxiomC{$
    \varHypo \cup \{(H,\statusUnknown)\},\,
    \varTracePrim,\,
    F_1 * \hformp{\formF^{\predP}_m}{\vec{u}}
    \entails
    F_2
  $}
  \def\extraVskip{3pt}
  \LeftLabel{\rulename{\ruleInduction}}
  \RightLabel{$\dagger_{(\ruleInduction)}$}
  \TrinaryInfC{$
    \varHypo \sepAnte\, \varTrace \sepAnte\,
    F_1 * \hformp{\predP}{\vec{u}}
    \entails
    F_2
  $}
\end{prooftree}
\end{minipage}
\\
\begin{minipage}{\textwidth}
\vspace{0.5em}
\begin{adjustwidth}{1.3em}{}
\begin{small}
Given
$H \triangleq F_1 * \hformp{\predP}{\vec{u}} \entails F_2$,~
$\varTracePrim = (\ruleInduction) \inserttrace \varTrace$,
and
$\dagger_{(\ruleInduction)}$:~
$\hformp{\predP}{\vec{u}} \triangleq
 \hformp{\formF^{\predP}_1}{\vec{u}} \mtor ...
 \mtor \hformp{\formF^{\predP}_m}{\vec{u}}$
\end{small}
\end{adjustwidth}
\end{minipage}
\end{tabular}
\end{small}

We show that if all of the entailments $F_1 *
\hformp{\formF^{\predP}_1}{\vec{u}} \entails F_2$,..., $F_1 *
\hformp{\formF^{\predP}_m}{\vec{u}} \entails F_2$ in the rule premise are valid,
then so is the entailment $F_1 * \hformp{\predP}{\vec{u}} \entails F_2$ in the
conclusion.

Indeed, consider an arbitrary model $s,h$ such that $s,h \satisfies F_1 *
\hformp{\predP}{\vec{u}}$. Side condition of the rule gives that
$\hformp{\predP}{\vec{u}} \triangleq \hformp{\formF^{\predP}_1}{\vec{u}} \mtor
... \mtor \hformp{\formF^{\predP}_m}{\vec{u}}$, i.e.,
$\hformp{\formF^{\predP}_1}{\vec{u}}$, ...,
$\hformp{\formF^{\predP}_m}{\vec{u}}$ are all definition cases of
$\hformp{\predP}{\vec{u}}$. Since $s,h \satisfies F_1 *
\hformp{\predP}{\vec{u}}$, it follows that $s,h \satisfies F_1 *
\hformp{\formF^{\predP}_i}{\vec{u}}$, for all $i\,{=\,}1...m$. On the other
hand, $F_1 * \hformp{\formF^{\predP}_1}{\vec{u}}$, ..., $F_1 *
\hformp{\formF^{\predP}_m}{\vec{u}}$ are antecedents of all entailments in this
rule's premises, and these entailments have the same consequent $F_2$.
Therefore, $s,h \satisfies F_2$. Since $s,h$ is chosen arbitrarily, it follows
that entailment in the rule's conclusion is valid. This confirms soundness of
the rule $\ruleInduction$. \qedhere

{\bf 8. Rule $\ruleHypo$:}

\begin{small}
\begin{tabular}{l}
\begin{minipage}{\textwidth}
\raggedleft
\begin{prooftree}
  \def\defaultHypSeparation{\hskip 2em}
  \def\ScoreOverhang{0em}
  \AxiomC{$
    \varHypo \cup \{(H, status)\},~
    (\ruleHypo) \inserttrace \varTrace,~
    F_4\theta \,{*}\, \Sigma' \,{\wedge}\, \Pi_1
    \entails
    F_2
  $}
  \def\extraVskip{3pt}
  \LeftLabel{\rulename{\ruleHypo}}
  \RightLabel{~\parbox{15em}{\rulesidecondright{
    \exists \theta{,} \Sigma'. (
    \Sigma_1 {\synequiv} \Sigma_3\theta {*} \Sigma'
    \mtand \Pi_1 {\Rightarrow} \Pi_3\theta)},\\
    \rulesidecondright{\dagger_{(\ruleHypo)}}
  }}
  \UnaryInfC{$
    \varHypo \,{\cup}\, \{(H \,{\triangleq}\, \Sigma_3 {\wedge} \Pi_3 {\entails} F_4, status)\},\,
    \varTrace,\,
    \Sigma_1 \,{\wedge}\, \Pi_1
    \,{\entails}\,
    F_2
  $}
\end{prooftree}
\end{minipage}
\\
\begin{minipage}{\textwidth}
\vspace{0.5em}
\begin{adjustwidth}{1em}{}
\begin{small}
\begin{tabular}{ll}
with $\dagger_{(\ruleHypo)}$:
$(status {=} \statusValid)$
&
$\mtor
\exists \varSort, u, \vec{v}, \Sigma''.
(\Sigma' \synequiv \hformntShort{\varSort}{u}{\vec{v}} * \Sigma'')$\\
{}&
$\mtor
\exists \varTrace_1, \varTrace_2.(
\varTrace \,{=}\, \varTrace_1 {\concattrace} [(\ruleStarData)] {\concattrace}
\varTrace_2
\mtand (\ruleInduction) \,{\nmembertrace}\, \varTrace_1
\mtand (\ruleInduction) \,{\membertrace}\, \varTrace_2)$.\\
\end{tabular}
\end{small}
\end{adjustwidth}
\end{minipage}
\end{tabular}
\end{small}

We show that if two entailments $F_4\theta \,{*}\, \Sigma' \,{\wedge}\, \Pi_1
\entails F_2$ and $\Sigma_3 \wedge \Pi_3 \entails F_4$ in the rule's premise are
valid, then so is the entailment in the rule's conclusion.

Indeed, the side condition $\Sigma_1 \synequiv \Sigma_3\theta * \Sigma'$ and
$\Pi_1 \Rightarrow \Pi_3\theta$ implies that $\Sigma_1 \wedge \Pi_1 \entails
\Sigma_3\theta * \Sigma' \wedge \Pi_3\theta \wedge \Pi_1$ is valid.

By applying Reynolds's substitution law~\cite{Reynolds08}, the hypothesis
$\Sigma_3 \wedge \Pi_3 \entails F_4$ implies that $\Sigma_3\theta \wedge
\Pi_3\theta \entails F_4\theta$ is valid. It follows that the following
entailment is also valid: $\Sigma_3\theta * \Sigma' \wedge \Pi_3\theta \wedge
\Pi_1 \entails F_4\theta * \Sigma' \wedge \Pi_1$.

We have shown that the two entailments $\Sigma_1 \wedge \Pi_1 \entails
\Sigma_3\theta * \Sigma' \wedge \Pi_3\theta \wedge \Pi_1$ and $\Sigma_3\theta *
\Sigma' \wedge \Pi_3\theta \wedge \Pi_1 \entails F_4\theta * \Sigma' \wedge
\Pi_1$ are valid. In addition, the rule's premise gives that $F_4\theta \,{*}\,
\Sigma' \,{\wedge}\, \Pi_1 \entails F_2$ is valid. It follows that the
entailment $\Sigma_1 \,{\wedge}\, \Pi_1 \,{\entails}\, F_2$ in the rule's
conclusion is valid as well. Therefore, the rule $\ruleHypo$ is correct.
\qedhere

\subsection{Soundness of the proof system}

We are now ready to prove soundness of our proof system, which is stated in
Theorem \ref{thm:Soundness}: {\em Given an entailment $E$, if the proof search
  procedure returns $\valValid$ when proving $E$, then $E$ is valid}.

Our soundness proof is as follow.

Suppose that during proving the entailment $E$, the proof search procedure
derives a proof tree $\varTree$ of $E$. If the rule $\ruleHypo$ ({\em apply
  hypothesis}) is not used in $\varTree$, then it is clear that induction
is not used in proof of $E$. Then, it is clear that $E$ is valid, since
soundness of our inference rules are proven in the previous section.

If the rule $\ruleHypo$ is used in $\varTree$, then statuses of the applied
hypotheses, at the moment they are used, can be either {\em valid}
($\statusValid$) or {\em unknown} ($\statusUnknown$). If statuses of all
the applied hypotheses in the proof tree $\varTree$ are valid
($\statusValid$), then it is clear that $E$ is valid, due to correctness of
our inference rules. If there exists some applied hypotheses whose statuses
are unknown ($\statusUnknown$), then these hypotheses may participate in a
mutual induction proof. We will show that the entailment $E$ is also valid
in this case.

Recall that hypotheses applied by the rule $\ruleHypo$ are either (i) induction
hypotheses recorded by the rule $\ruleInduction$, or (ii) hypotheses derived by
other rules during proof search. Therefore, in the proof tree $\varTree$, both
the induction hypotheses and other hypotheses can participate in a mutual
induction proof. We will transform the proof tree $\varTree$ into a new tree
$\varTree'$ in which the mutual induction proof involves only the induction
hypotheses recorded by the rule $\ruleInduction$. This can be done by modifying
the rule $\ruleHypo$ to put the used hypothesis, which is not derived by the
rule $\ruleInduction$, into premises of the rule.

In particular, suppose the (simplified) rule $\ruleHypo$ which applies a
hypothesis $\Sigma_3 \wedge \Pi_3 \entails F_4$, (not derived by the rule
$\ruleInduction$), into proving a goal entailment $\Sigma_1 \wedge \Pi_1
\entails F_2$ as follow:

\begin{small}
\begin{tabular}{l}
\begin{minipage}{\textwidth}
\raggedleft
\begin{prooftree}
  \def\defaultHypSeparation{\hskip 2em}
  \def\ScoreOverhang{0em}
  \AxiomC{$
    F_4\theta * \Sigma'
    \entails
    F_2
  $}
  \def\extraVskip{3pt}
  \LeftLabel{\rulename{\ruleHypo}}
  \RightLabel{~\parbox{30em}{
  \rulesidecondright{
    \text{apply hypothesis:~} \Sigma_3 \wedge \Pi_3 \entails F_4,
    \text{\,(which is not derived by rule $\ruleInduction$)}}
    \\
  \rulesidecondright{
    \exists\, \theta, \Sigma'. (
    \Sigma_1 \,{\equiv}\, \Sigma_3\theta \,{*}\, \Sigma'
    \,\mtand\, \Pi_1 \,{\Rightarrow}\, \Pi_3\theta)}
  }}
  \UnaryInfC{$
    \Sigma_1 \,{\wedge}\, \Pi_1
    \,{\entails}\,
    F_2
  $}
\end{prooftree}
\end{minipage}
\end{tabular}
\end{small}
\\[0.5em]

We modify the rule $\ruleHypo$ so that the hypothesis $\Sigma_3 \wedge \Pi_3
\entails F_4$ appears in the premises as follows:

\begin{small}
\begin{tabular}{l}
\begin{minipage}{\textwidth}
\raggedleft
\begin{prooftree}
  \def\defaultHypSeparation{\hskip 1.5em}
  \def\ScoreOverhang{0em}
  \AxiomC{$
    \Sigma_3 \wedge \Pi_3 \entails F_4
  $}
  \AxiomC{$
    F_4\theta * \Sigma'
    \entails
    F_2
  $}
  \def\extraVskip{3pt}
  \LeftLabel{\rulename{\text{modified}~\ruleHypo}}
  \RightLabel{~\parbox{30em}{
  \rulesidecondright{
    \Sigma_3 \wedge \Pi_3 \entails F_4
    \text{~is not derived by rule $\ruleInduction$}}
    \\
  \rulesidecondright{
    \exists\, \theta, \Sigma'. (
    \Sigma_1 \,{\equiv}\, \Sigma_3\theta \,{*}\, \Sigma'
    \,\mtand\, \Pi_1 \,{\Rightarrow}\, \Pi_3\theta)}
  }}
  \BinaryInfC{$
    \Sigma_1 \,{\wedge}\, \Pi_1
    \,{\entails}\,
    F_2
  $}
\end{prooftree}
\end{minipage}
\end{tabular}
\end{small}
\\[0.5em]

\begin{tabular}{ll}
\begin{minipage}{0.45\textwidth}
\begin{figure}[H]
\begin{tikzpicture}[
  ->,>=stealth',shorten >=1pt,auto,node distance=2.4cm,thick,
  main node/.style={circle,draw,font=\sffamily},
  label/.style={draw=none},
  hidden node/.style={font=\sffamily},
  hidden point/.style={width=1pt, inner sep=0pt}
  ]

  \node[main node] (E) {};
  \node[label] [below right =0cm and -0.3cm of E] (L) {{
      \scriptsize $\Sigma_1 {\wedge} \Pi_1 {\entails} F_2, (\ruleHypo)$}};

  \node[main node] (E1) [above right =0.8cm and 0.5cm of E] {};
  \node[label] [below right =-0.3cm and -2cm of E1] (L) {{
      \scriptsize $F_4\theta {*} \Sigma' {\entails} F_2$}};
  \node[label] [above right =0.4cm and -0.5cm of E1] (L) {{
      \scriptsize $\mathcal{T}_2$}};

  \node[main node] (H) [left =0cm and 1.6cm of E] {};
  \node[label] [below right =0.05cm and -1.8cm of H] (L) {{
      \scriptsize $\Sigma_3 {\wedge} \Pi_3 {\entails} F_4$}};
  \node[label] [above right =0.3cm and -0.6cm of H] (L) {{
      \scriptsize $\mathcal{T}_1$}};

  \node[hidden node] (R) [below left=1.2cm and 1.2cm of E] {};
  \node[label] [below = -0.2cm of R] (L) {{\scriptsize $E$}};
  \node[label] [below = 0.1cm of R] (L) {{\scriptsize $Root$}};

  \coordinate  [above left=1cm and 0.65cm of H] (HLeft);
  \coordinate  [above right=1cm and 0.2cm of H] (HRight);

  \coordinate  [above left=1.1cm and 0.55cm of E1] (E1Left);
  \coordinate  [above right=1.1cm and 0.6cm of E1] (E1Right);

  \draw [dotted,-] (H) to[out=160, in=-130] (HLeft);
  \draw [dotted,-] (H) to[out=30, in=-30] (HRight);
  \draw [dotted,-] (HLeft) to[out=20, in=160](HRight);

  \draw [dotted,-] (E1) to[out=160, in=-130] (E1Left);
  \draw [dotted,-] (E1) to[out=30, in=-30] (E1Right);
  \draw [dotted,-] (E1Left) to[out=20, in=160] (E1Right);

  \draw [rounded corners,dotted] (H) -- node{{\scriptsize apply hypo}} (E);
  \draw [rounded corners,dashed] (R) -- (E);
  \draw [rounded corners] (E) -- (E1);
  \draw [rounded corners,dashed] (R) -- (H);
\end{tikzpicture}
\caption{Original proof tree $\varTree$,
where $\Sigma_3 \wedge \Pi_3 \vdash F_4$
is not derived by the induction rule $\ruleInduction$}
\label{fig:ProofTree}
\end{figure}
\end{minipage}
&
\begin{minipage}{0.6\textwidth}
\begin{figure}[H]
\begin{tikzpicture}[
  ->,>=stealth',shorten >=1pt,auto,node distance=2.4cm,thick,
  main node/.style={circle,draw,font=\sffamily},
  label/.style={draw=none},
  hidden node/.style={font=\sffamily}]

  \node[main node] (E) {};
  \node[label] [below right =0.05cm and -0.5cm of E] (L) {{
      \scriptsize $\Sigma_1 {\wedge} \Pi_1 {\entails} F_2$}};

  \node[main node] (E1) [above right =0.8cm and 1cm of E] {};
  \node[label] [below right =0.1cm and -0.3cm of E1] (L) {{
      \scriptsize $F_4\theta {*} \Sigma' {\entails} F_2$}};
  \node[label] [above right =0.4cm and -0.5cm of E1] (L) {{
      \scriptsize $\mathcal{T}_2$}};

  \node[main node] (H) [left =0cm and 2.4cm of E] {};
  \node[label] [below right =-0.2cm and 0.05cm of H] (L) {{
      \scriptsize $\Sigma_3 {\wedge} \Pi_3 {\entails} F_4$}};
  \node[label] [above right =0.3cm and -0.6cm of H] (L) {{
      \scriptsize $\mathcal{T}_1$}};

  \node[main node] (H1) [above left =0.8cm and 0.8cm of E] {};
  \node[label] [below right =-0.2cm and 0.05cm of H1] (L) {{
      \scriptsize $\Sigma_3 {\wedge} \Pi_3 {\entails} F_4$}};
  \node[label] [above right =0.55cm and -1cm of H1] (L) {{
      \scriptsize inserted}};
  \node[label] [above left =0.15cm and -0.55cm of H1] (L) {{
      \scriptsize tree $\mathcal{T}'_1$}};

  \node[hidden node] (R) [below left=1.2cm and 1.2cm of E] {};
  \node[label] [below = -0.2cm of R] (L) {{\scriptsize $E$}};
  \node[label] [below = 0.1cm of R] (L) {{\scriptsize $Root$}};

  \coordinate  [above left=1cm and 0.65cm of H] (HLeft);
  \coordinate  [above right=1cm and 0.2cm of H] (HRight);

  \coordinate  [above left=1cm and 0.65cm of H1] (H1Left);
  \coordinate  [above right=1cm and 0.2cm of H1] (H1Right);

  \coordinate  [above left=1.1cm and 0.55cm of E1] (E1Left);
  \coordinate  [above right=1.1cm and 0.6cm of E1] (E1Right);

  \draw [dotted,-] (H) to[out=160, in=-130] (HLeft);
  \draw [dotted,-] (H) to[out=30, in=-30] (HRight);
  \draw [dotted,-] (HLeft) to[out=20, in=160](HRight);

  \draw [dotted,-] (H1) to[out=160, in=-130] (H1Left);
  \draw [dotted,-] (H1) to[out=30, in=-30] (H1Right);
  \draw [dotted,-] (H1Left) to[out=20, in=160] (H1Right);

  \draw [dotted,-] (E1) to[out=160, in=-130] (E1Left);
  \draw [dotted,-] (E1) to[out=30, in=-30] (E1Right);
  \draw [dotted,-] (E1Left) to[out=20, in=160] (E1Right);

  \draw [rounded corners,dashed] (R) -- (E);
  \draw [rounded corners] (E) -- (H1);
  \draw [rounded corners] (E) -- (E1);
  \draw [rounded corners,dashed] (R) -- (H);
\end{tikzpicture}
\caption{Transformed proof tree $\varTree'$}
\label{fig:ProofTreeTwo}
\end{figure}
\end{minipage}
\end{tabular}

By modifying the rule $\ruleHypo$, we can transform the proof tree $\varTree$
into a proof tree $\varTree'$, where every node of the tree $\varTree$ which
applies a hypothesis (not derived by the rule $\ruleInduction$) to prove an
entailment is replaced by a new node that contains not only the target
entailment, but also full proof tree of the applied hypothesis. Since this
transformation is performed only on hypotheses not derived by the rule
$\ruleInduction$, it follows that in the new proof tree $\varTree'$, only
induction hypotheses recorded by the rule $\ruleInduction$ participating into the
mutual induction proof. We have shown earlier in the discussion about the rule
$\ruleHypo$ that its side condition about model decreasing maps to the
well-founded relation of our mutual induction principle.
Therefore, the mutual induction principle is ensured by this
new proof tree. This implies that all induction hypotheses (recorded by the rule
$\ruleInduction$) are valid. Therefore, the original entailment is also valid.
\qedhere

\end{document}